\documentclass[aps,prx,twocolumn,superscriptaddress,showpacs,showkeys, 10pt, longbibliography]{revtex4-2}
\usepackage{natbib}
\usepackage{graphicx}
\usepackage{amsmath}
\usepackage{amssymb}
\usepackage{bm}
\usepackage{color,soul}
\usepackage{hyperref}
\usepackage{float}
\usepackage{csquotes}
\usepackage{color,soul}
\usepackage{lipsum}
\hypersetup{colorlinks=true, urlcolor=blue, linkcolor = blue, citecolor=blue}
\usepackage{braket}
\usepackage[utf8]{inputenc}
\usepackage{enumitem}
\usepackage{threeparttable}
\usepackage{longtable}
\usepackage{threeparttablex}
\usepackage{multirow}
\usepackage{ulem}

\setlist[enumerate,1]{label=\arabic*.}              

\setlist[enumerate,2]{label*=\arabic*.}             
\setlist[enumerate,3]{label*=\arabic*.}             

\begin{document}

\title{Tunable Hybridization of Whispering Gallery Modes in High-Index Photonic Molecules}

\author{Ayusmin Panda}
\email{pandaayusmin97@gmail.com}
\affiliation{Condensed Matter Theory and Computational Lab, Department of Physics, IIT Madras, India}
\affiliation{Center for Atomistic Modeling and Materials Design, IIT Madras, India} 
\author{Birabar Ranjit Kumar Nanda}
\email{nandab@iitm.ac.in}
\affiliation{Condensed Matter Theory and Computational Lab, Department of Physics, IIT Madras, India}
\affiliation{Center for Atomistic Modeling and Materials Design, IIT Madras, India} 

\begin{abstract}
Photonic molecules formed by coupled dielectric microresonators provide a platform for controlling collective optical modes, yet the interplay between molecular topology, resonator detuning, and dielectric environment in determining their spectral response remains insufficiently quantified. Here, we systematically quantify the evolution of Whispering Gallery Mode (WGM) hybridization in high-index TiO$_2$ and perovskite coated TiO$_2$ microresonators with molecular topology, and establish the corresponding effective inter-cavity coupling using a combined numerical and analytical analysis. We further show that perovskite-shell incorporation provides a diameter and thickness-dependent modification of the effective coupling, producing a non-monotonic response rather than a uniform suppression of coupling. Despite the resulting closely spaced hybridized resonances in the high-index perovskite environment, the supermodes remain spectrally resolvable, with $Q$-factor of order \(10^3\) for representative modes. The observed spectral splitting is quantitatively related to the effective inter-cavity coupling strengths, revealing a non-monotonic coupling response to perovskite-shell incorporation that depends on resonator size and shell thickness. These results establish molecular topology and dielectric environment as distinct parameters for controlling WGM hybridization, inter-cavity coupling, and spectral response, providing a basis for the design of coupled resonator systems for applications in optical sensing, filtering, switching, and other resonant photonic devices. \\
\end{abstract}


\maketitle

\section{Introduction}
Optical microresonators supporting whispering gallery modes (WGMs) have emerged as indispensable building blocks in modern photonics owing to their ability to confine light efficiently within compact dielectric cavities while sustaining exceptionally high quality ($Q$) factors. These characteristics enable strong light–matter interactions, prolonged photon lifetimes, and high spectral selectivity, making WGM resonators attractive for a wide range of applications including ultrasensitive refractive index sensing, single particle detection, narrow-linewidth microlasers, nonlinear frequency conversion, and optical filtering \cite{Foreman2015, Vollmer2008}. The sharpness of a WGM resonance is determined by its $Q$-factor, which quantifies the balance between energy storage and optical loss within the cavity. High-$Q$ resonances promote enhanced field buildup and prolonged optical confinement, thereby improving sensitivity to external perturbations and enabling precise spectral control. These unique properties have established WGM microresonators as a versatile platform for integrated photonics and emerging optoelectronic systems, including perovskite-based photonic devices \cite{Arnold2003, Boriskina2006, Panda2026}.

When two or more microresonators are placed in close proximity, their evanescent WGM fields overlap, giving rise to hybridized optical states known as photonic molecules \cite{Liao2020, Fuller1991}. In direct analogy to electronic molecular orbitals, the WGMs of isolated spheres split into bonding and antibonding supermodes whose frequency separation depends on the strength of inter-cavity coupling \cite{Boriskina2006}. The degree of mode splitting is governed by the center-to-center separation, mode symmetry, refractive index contrast, and the topology of the resonator system \cite{Arnold2001, Kanev2006}. Previous studies have shown that photonic molecules formed from polymer or silica spheres exhibit rich spectral behavior, including multi-peak structures, mode degeneracies, and discrete ``spectral fingerprints" that uniquely encode the geometry of the coupled system \cite{Li2017, Li2015}. These features have motivated their use in photonic circuits \cite{Ren2025, Zhang2019, BORISKINA2013}, reconfigurable networks \cite{Ahn2012, Ferraro2017}, cluster-based sensors \cite{Zhang2025, Hara2005}, and optically bound microstructures \cite{Helgason2021, BORISKINA2013375}.

The hybridization of WGMs enhances the sensitivity of coupled resonators to environmental perturbations, enabling detection schemes capable of distinguishing bulk refractive-index variations from surface interactions, as well as specific and non-specific binding events \cite{Astratov:13, Arnold2009}. Moreover, mode splitting in coupled resonators provides a robust mechanism for high-resolution detection and nanoparticle sensing \cite{VollmerYang2012}. Beyond sensing, these structures enable controlled engineering of the photonic density of states, supporting applications in low-threshold lasing, CROWs - coupled resonator optical waveguides with tunable delay, and directional light transport \cite{Moller2007, Astratov2004, Yariv:99, Franchimon2013}. Their ability to maintain well-defined optical modes while offering {multiple degrees of freedom for interaction makes photonic molecules a versatile platform for next-generation integrated photonic and optoelectronic devices.

 \begin{figure*}
     \centering
     \includegraphics[scale=0.38]{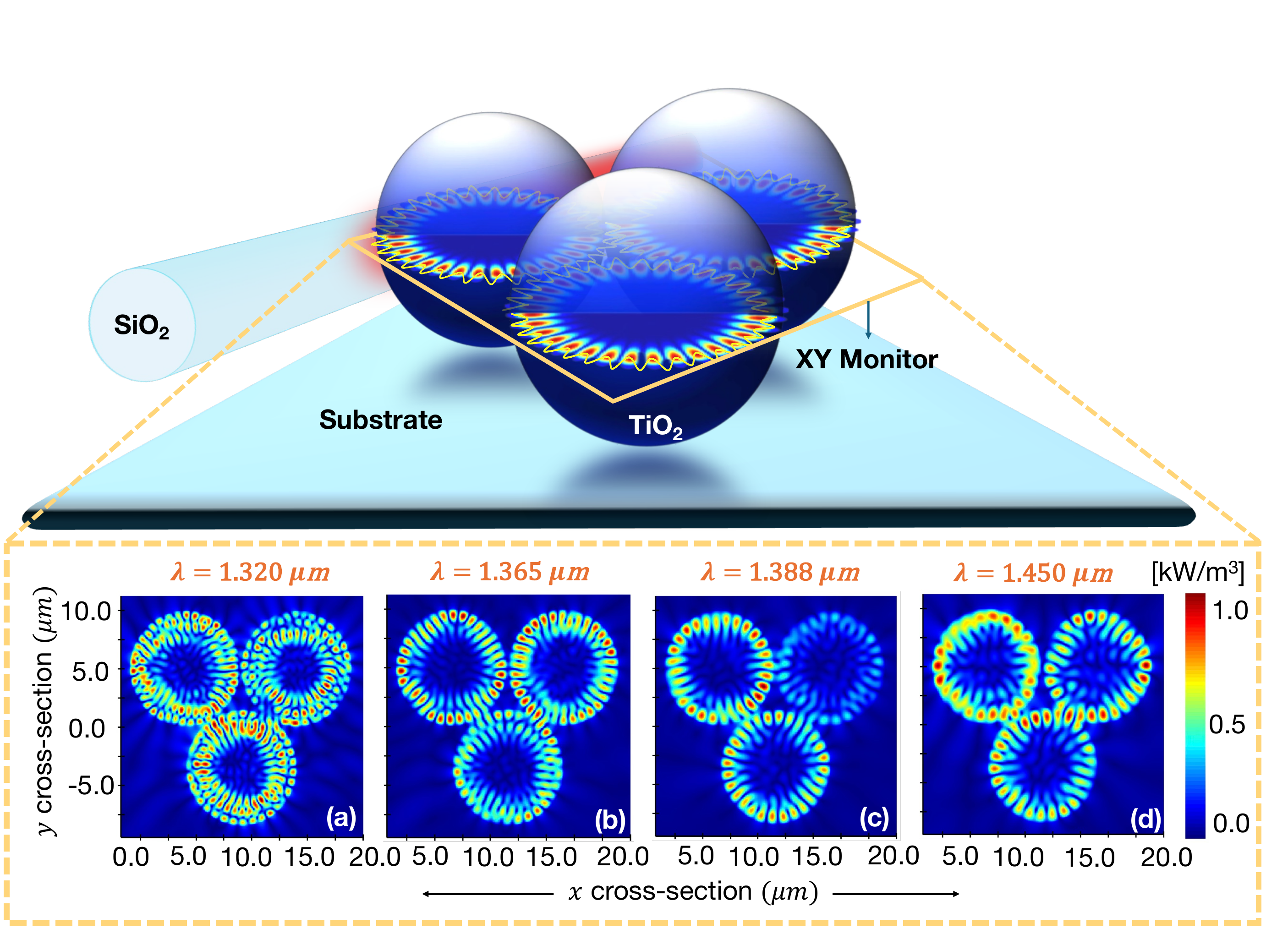}
     \caption{ Schematic illustration of coupled whispering gallery mode resonators and spatial field distributions of hybridized supermodes. (Top) Three-sphere photonic molecule configuration evanescently excited using tapered SiO$_2$ ﬁber, enabling inter-cavity coupling. The lower panels (a–d) shows the XY cross-sectional view of the variation in the electric field distribution profile as calculated using numerical simulations for the three-molecule system, for various emission wavelength. This illustrates the propagation of light through evolution of hybridized supermodes at various wavelengths.}
     \label{fig:1}
 \end{figure*}

The spectral response of a photonic molecule, including its resonance multiplicity, mode spacing, linewidth, and field distribution, is fundamentally governed by the geometry, refractive index contrast, and optical coupling between its constituent resonators \cite{Li2013}. Although these relationships have been extensively investigated in silica and polymer-based photonic molecules, comparatively little attention has been devoted to high-index dielectric resonators, despite their ability to provide stronger optical confinement and enhanced evanescent coupling in the visible and near-infrared spectral regions. In particular, a systematic understanding of how cavity topology and dielectric engineering jointly influence optical supermode formation and spectral organization in high-index photonic molecules remains limited.

High-index TiO$_2$ microresonators provide an attractive platform to address this challenge owing to their enhanced boundary field localization, which strengthens inter-cavity interactions and increases sensitivity to dielectric perturbations \cite{Teraoka2006, VollmerYang2012}. Furthermore, introducing a perovskite shell modifies the local dielectric environment, redistributes the WGM field towards the shell-medium interface, and introduces an additional degree of freedom for tuning inter-cavity coupling without altering the molecular topology \cite{Teraoka:07}. Such dielectric core-shell engineering offers the possibility of independently controlling mode splitting, spectral compression, and modal redistribution, thereby extending the design space beyond conventional geometry-driven approaches.

In this work, we establish a unified framework for understanding optical supermode formation in high-index photonic molecules through the combined influence of cavity topology and dielectric-shell engineering. Using finite-difference time-domain simulations together with a single-mode tight-binding model, we systematically investigate pristine TiO$_2$ and CsPbBr$_3$-coated TiO$_2$ core–shell photonic molecules with linear and compact geometries. We demonstrate that each molecular topology exhibits a distinct spectral fingerprint governed by the number, symmetry, and connectivity of the coupled resonators, while the introduction of a perovskite shell provides an independent mechanism for tailoring inter-cavity coupling through controlled dielectric perturbation. Consequently, dielectric-shell engineering provides an additional degree of freedom for tailoring the spectral distribution of hybridized supermodes. Depending on the molecular topology and shell thickness, the resulting dielectric perturbation modifies the relative detuning and spectral separation of the hybridized modes, enabling reduced supermode spectral separation while retaining spectrally resolvable resonances with $Q$-factors of order \(10^3\) for representative modes. Comparison with the full-wave simulations supports the coupled-mode interpretation of the observed supermode splitting. This provides physical insight into the evolution of hybridized optical supermodes and establishes general design principles for engineering coupled microresonator systems for integrated photonics, spectral filtering, and optical sensing. The proposed framework is general and can be extended to a broad class of high-index coupled microresonator systems beyond the TiO$_2$/perovskite platform.

\section{Topology-Dependent Optical Supermodes in High-Index Photonic Molecules}\label{sec:numerical simulations}
We have used a simplified 2D model for studying the underlying physical mechanisms of optically coupled micropsheres. The WGMs in microspheres are characterized by radial ($n$), angular ($l$) and azimuthal ($m$) mode numbers, which describe the field variation along the radial, polar and circumferential, directions, respectively \cite{Teraoka2006, Teraoka2006, Panda2024}. The fundamental mode corresponds to ($n$ = 1) and ($l$ = $m$), indicating a single radial maximum with strongest confinement near the equatorial plane. Further in tapered fiber coupling-based experiments, the excitation of WGMs is governed by the contact position and orientation of the taper, typically favoring modes confined to the equatorial plane \cite{Arnold2001, Griffel1996}. Therefore, as schematically shown in Figure~\ref{fig:1}, photonic molecules assembled on substrates can be effectively treated within a quasi-2D framework, where spheres are modeled as coupled circular resonators and the tapered fiber as a nearby waveguide. This reduced representation is used to capture the essential in-plane coupling and hybridization of the selected WGM family, rather than to reproduce all three-dimensional modal degrees of freedom of an individual microsphere. Figure 1 (bottom inset) shows the formation of hybridized supermodes in a three-cavity photonic molecule as an example, where each wavelength selectively excites a distinct eigen state of the coupled system. The different wavelengths correspond to distinct supermode eigen states with different phase and spatial distributions across the coupled cavities, ranging from delocalized to increasingly localized field configurations \cite{Li2017}. This evolution directly reflects the wavelength-dependent redistribution of optical energy governed by inter-cavity coupling.

\begin{figure}
     \centering
     \includegraphics[scale=0.33]{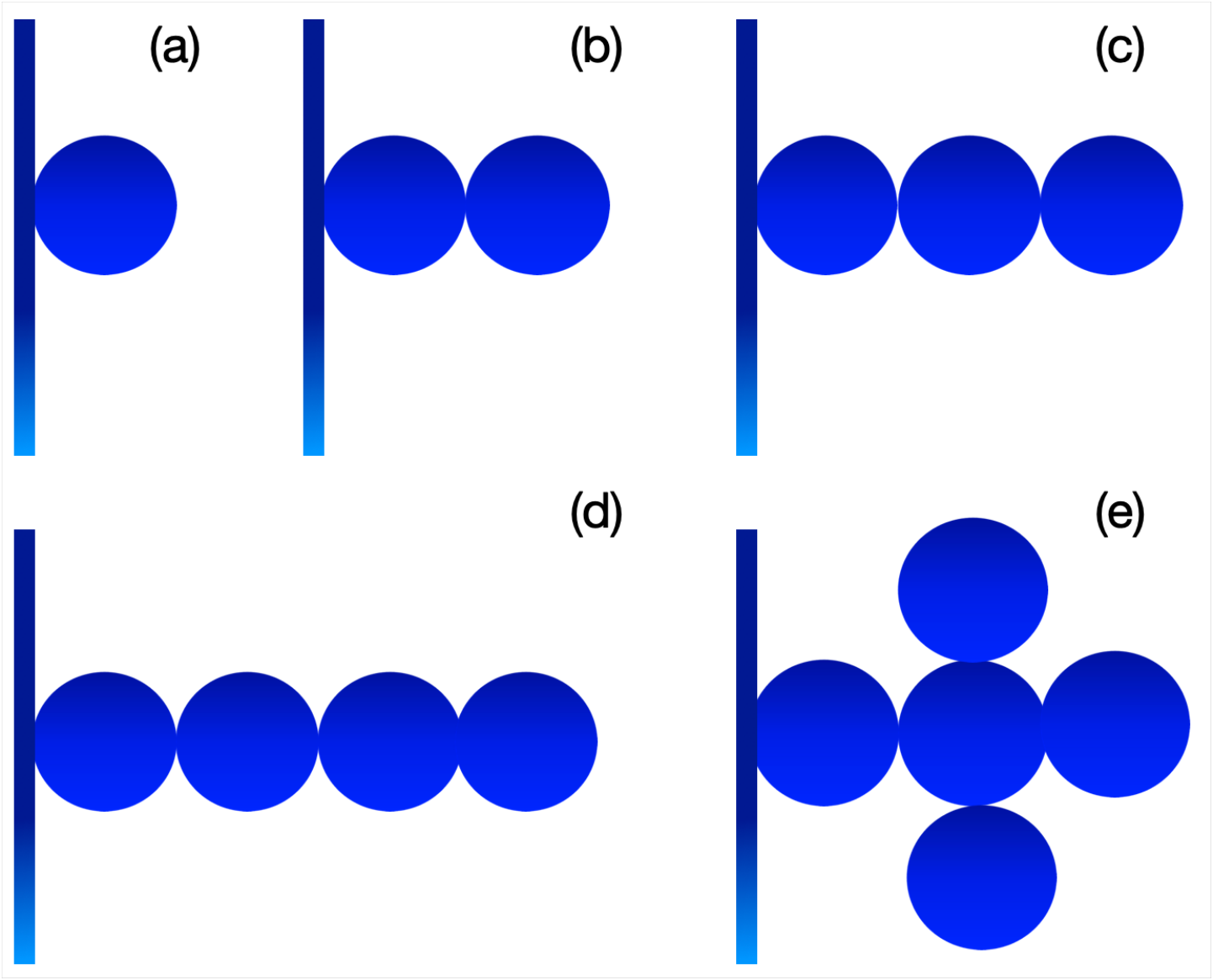}
     \caption{Schematic illustration of the various photonic molecule configuration withside coupling by tapered ﬁber: (a) single sphere, linear chains of (b) two, (c) three, (d) four circular resonators and (e) five atom cross circular resonators.}
     \label{fig:2}
\end{figure}

To elucidate optical coupling in photonic molecules, we extend the analysis from bispheres to linear chains and clusters of pristine and perovskite-coated TiO$_2$ microspheres (Figure~\ref{fig:2}). Increasing the number of spheres leads to systematic splitting of isolated WGMs into multiple hybridized supermodes, with topology and symmetry defining distinct spectral structures \cite{Li2017, Li2015}. While the magnitude of splitting depends on sphere diameter and refractive index contrast, the topology-dependent spectral organization remains identifiable, forming characteristic optical fingerprints.

WGM hybridization is governed by the interplay between coupling $g$, arising from evanescent field overlap, and detuning $\Delta$, set by intrinsic resonance differences. Strong coupling ($g$ $\gg$ $\Delta$) yields symmetric delocalized modes, whereas ($g$ $\sim$ $\Delta$) and ($g$ $\ll$ $\Delta$) lead to asymmetric and localized states. Here, geometry controls $g$, while perovskite coating modifies the effective detuning and coupling through re-distribution of the WGM field and changes in inter-cavity overlap.

 \begin{figure*}
     \centering
     \includegraphics[scale=0.28]{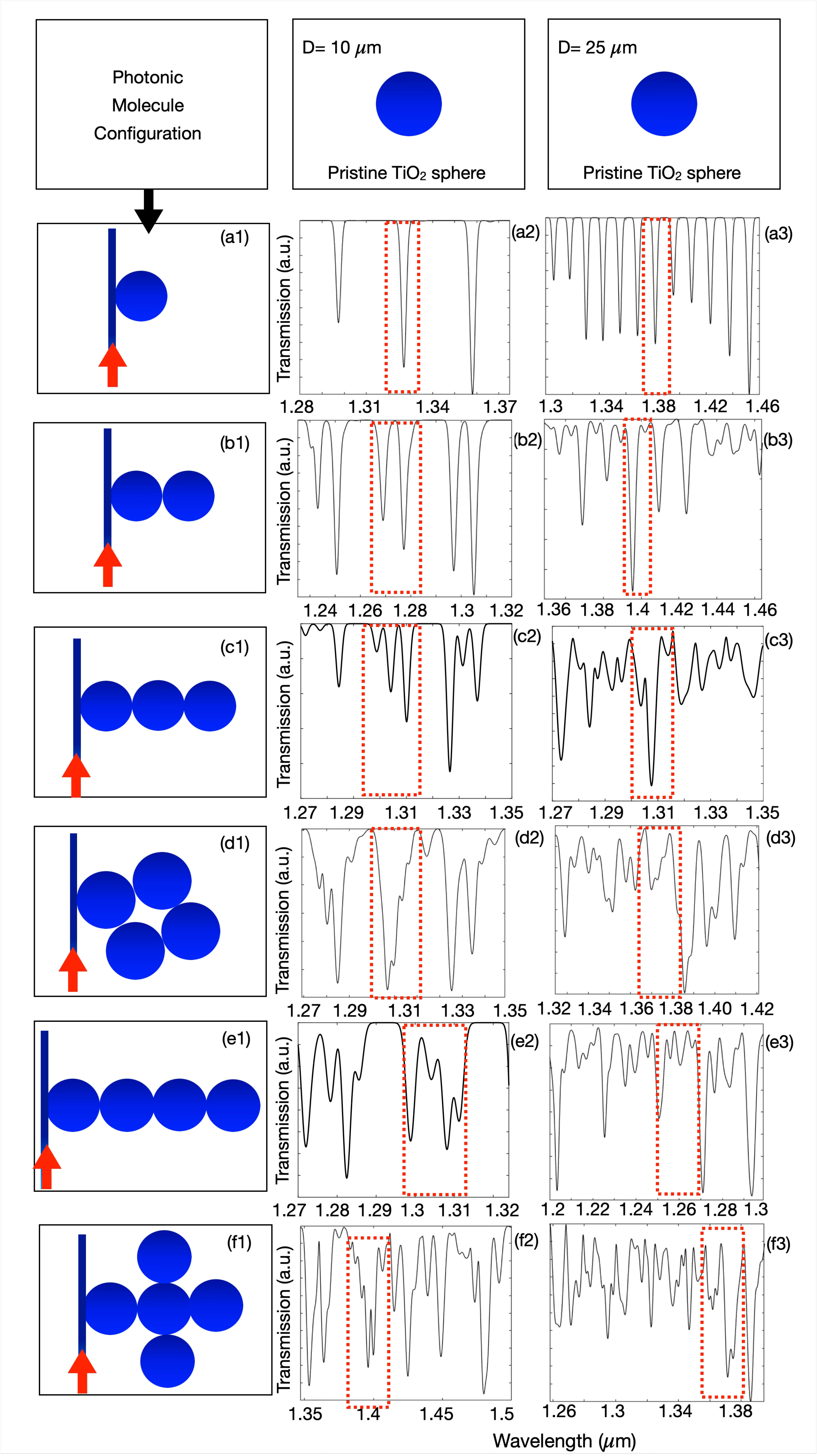}
     \caption{Simulated transmission spectra of hybridization of WGM in TiO$_2$ photonic molecules. Each uncoupled WGM resonance (a1 and a2) forms a set of supermodes with splitting pattern (b2-f2) for cavity diameter $D$=10 $\mu$m, and (b3-f3) for cavity diameter $D$=25 $\mu$m, with representative given molecular configurations illustrated in (b1-f1), respectively. Molecular configurations including: (a1) single cavity, (b1) bi-cavity, linear chain of (c1) 3 circular resonators and (e1) 4 circular resonators, and planar (d1) quadrumer, and (f1) 5 cross resonator.}
     \label{fig:3}
 \end{figure*}
 
\subsection{Pristine dielectric spheres}
The isolated TiO$_2$ microcavity exhibits a series of periodic whispering gallery mode resonances that serve as the parent eigenfrequencies ($\lambda_0$) for all coupled configurations. As shown in Figure~\ref{fig:3}a2 (10 $\mu$m dia) and Figure~\ref{fig:3}a3 (25 $\mu$m dia), the free spectral range (FSR) decreases with increasing cavity diameter owing to the longer optical path length within the resonator. These uncoupled resonances provide the reference against which the hybridization-induced spectral evolution in photonic molecules is quantified.

When two resonators are brought into contact, coherent evanescent coupling lifts the degeneracy of the isolated WGM, causing each parent resonance to split into two hybridized supermodes symmetrically distributed about $\lambda_0$. The observed splitting remains significantly smaller than the intrinsic FSR of the isolated cavity, ensuring that hybridization occurs within the same radial WGM manifold rather than through coupling between different radial mode families. Furthermore, the resonance linewidths remain substantially smaller than the mode splitting owing to the moderate cavity quality factors ($Q$ $\approx$ 10$^3$), allowing the hybridized states to be clearly resolved.

The evolution of the transmission spectra with increasing resonator number reveals a clear dependence on molecular topology. While the transmission spectra in Figure~\ref{fig:3} reveal the qualitative evolution of hybridized WGMs, a quantitative description of topology-dependent spectral organization requires analysis of the relative positions of the supermodes with respect to the 
parent cavity resonance. For each resonator configuration, an isolated-cavity resonance `$\lambda_o$' was selected as the spectral reference. Upon coupling identical resonators to form a photonic molecule, the resonances appearing in the spectral vicinity of this reference were considered as the corresponding hybridized modes and were characterized by their normalized detuning, defined as $\Delta_i$ = $(\lambda_i -\lambda_o)/\lambda_o$, where $\lambda_i$ denotes the wavelength of the $i$-th resolved hybridized resonance. Accordingly, the analysis does not require an explicit assignment of the radial or azimuthal WGM indices; instead, the spectral redistribution induced by resonator coupling is evaluated relative to the isolated-cavity reference resonance for each configuration. Therefore, Figure~\ref{fig:4} presents the normalized detuning of the hybridized modes for the various pristine TiO$_2$ photonic molecule configurations. Representing the spectra in terms of normalized detuning removes the influence of the absolute resonance wavelength and enables direct comparison of the spectral organization across different molecular topologies. For linear chains, the number of supermodes equals the number of constituent resonators, and the splitting follows an almost uniform progression. In the $N$ = 3 case, the central mode is slightly red-shifted relative to the uncoupled resonance, while adjacent modes are nearly equally spaced. This behavior extends to $N$ = 4, where four supermodes appear with equivalent separations \cite{Li2017}. However as shown in Figure~\ref{fig:A1}, for larger resonators with $N$ = 5 and $N$ = 6, three dominant spectral features are observed, corresponding to grouped eigen states within the same spectral envelope. This occurs due to near degeneracy of closely spaced modes, limiting the excitation of higher-order states, thereby leading to spectral merging. The key distinction between the two resonator molecule system is the position of central supermode. For 6-resonator system, the central supermode overlaps more with the uncoupled resonance, than the 5-resonator system. 

  \begin{figure}
     \centering
     \includegraphics[scale=0.21]{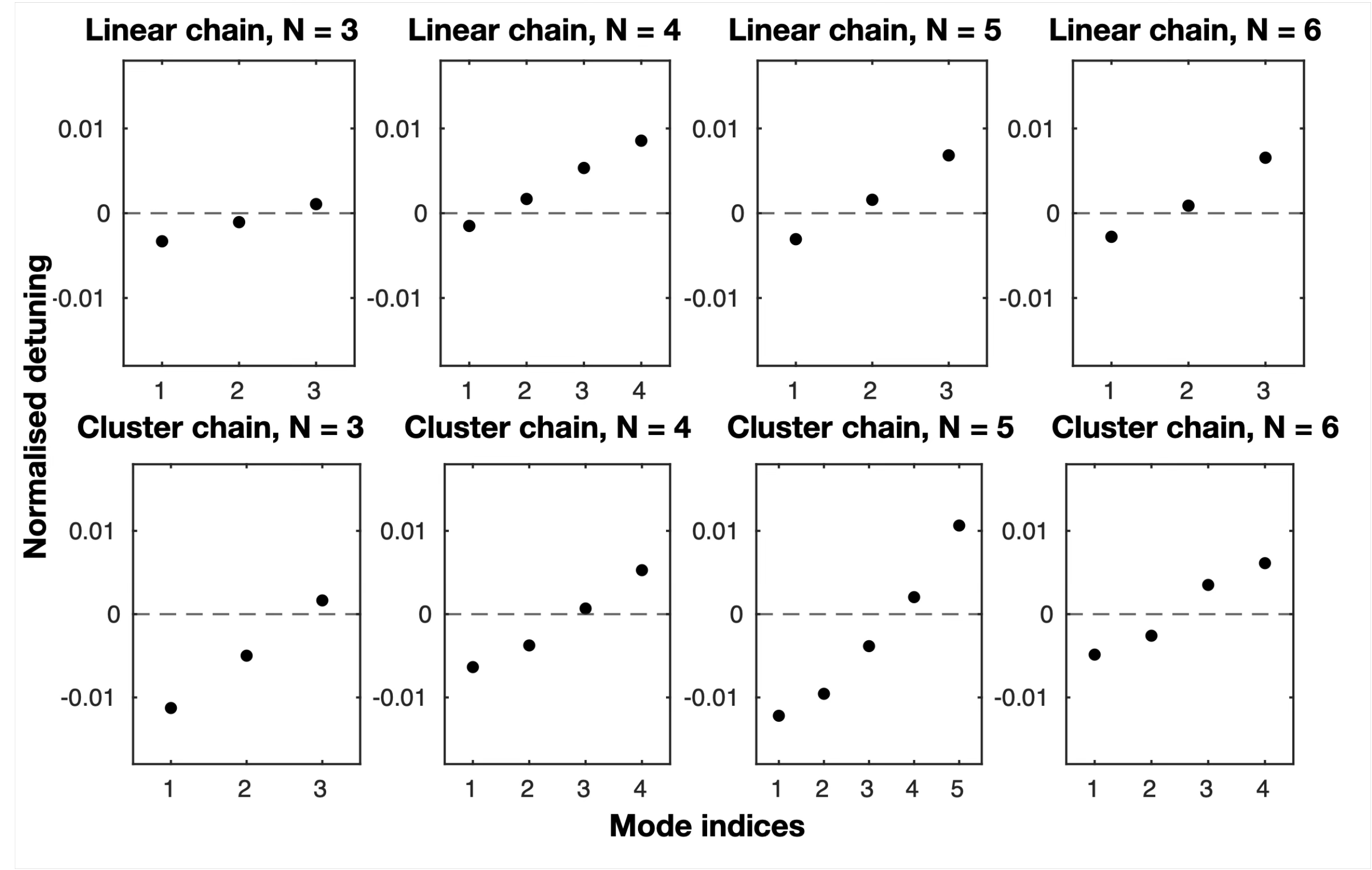}
     \caption{Normalized detuning ($\Delta$) of hybridized supermodes relative to the corresponding isolated-cavity parent resonance for linear and compact TiO$_2$ photonic molecular systems with $D$=10 $\mu$m. Linear chains show equi-spaced splitting that scales with the number of resonators, while compact clusters exhibit broadened and uneven spectral distributions due to multi-directional coupling. The variation in mode count and spacing serves as a characteristic spectral fingerprint of each configuration.}
     \label{fig:4}
 \end{figure}

In contrast, compact cluster configurations exhibit detuning patterns that deviate from uniform splitting due to the presence of multiple, non-equivalent coupling pathways. Even for $N$ = 3, the spacing between modes becomes uneven, indicating departure from the linear configuration behavior. For $N$ = 4 clusters, the splitting is no longer uniform, and the inner modes experiences a larger shift in detuning wavelength compared to linear chains, accompanied by unequal separations between adjacent modes. As the number of resonators increases, these deviations become more evident. For $N$ = 5 clusters, distinct five supermodes are visible in the set of two and three, about the uncoupled resonance mode. The splitting between the inner supermodes is smaller than the outer supermodes. For $N$ = 6, partial degeneracy emerges, leading to fewer distinct observable modes than the total number of resonators, along with uneven spacing between the remaining modes.

These above-mentioned features such as non-uniform splitting, variation in central mode shift, and the presence of degeneracy constitute characteristic spectral fingerprints that distinguish compact clusters from linear chains. Importantly, while the absolute spectral positions and free spectral range may vary with the cavity size, the qualitative detuning distribution remains governed by the connectivity and symmetry of the molecular configuration. This confirms that the spectral organization of hybridized supermodes is dictated primarily by topology, enabling identification of photonic molecule geometry directly from their spectral response. 

\begin{figure*}
     \centering
     \includegraphics[scale=0.3]{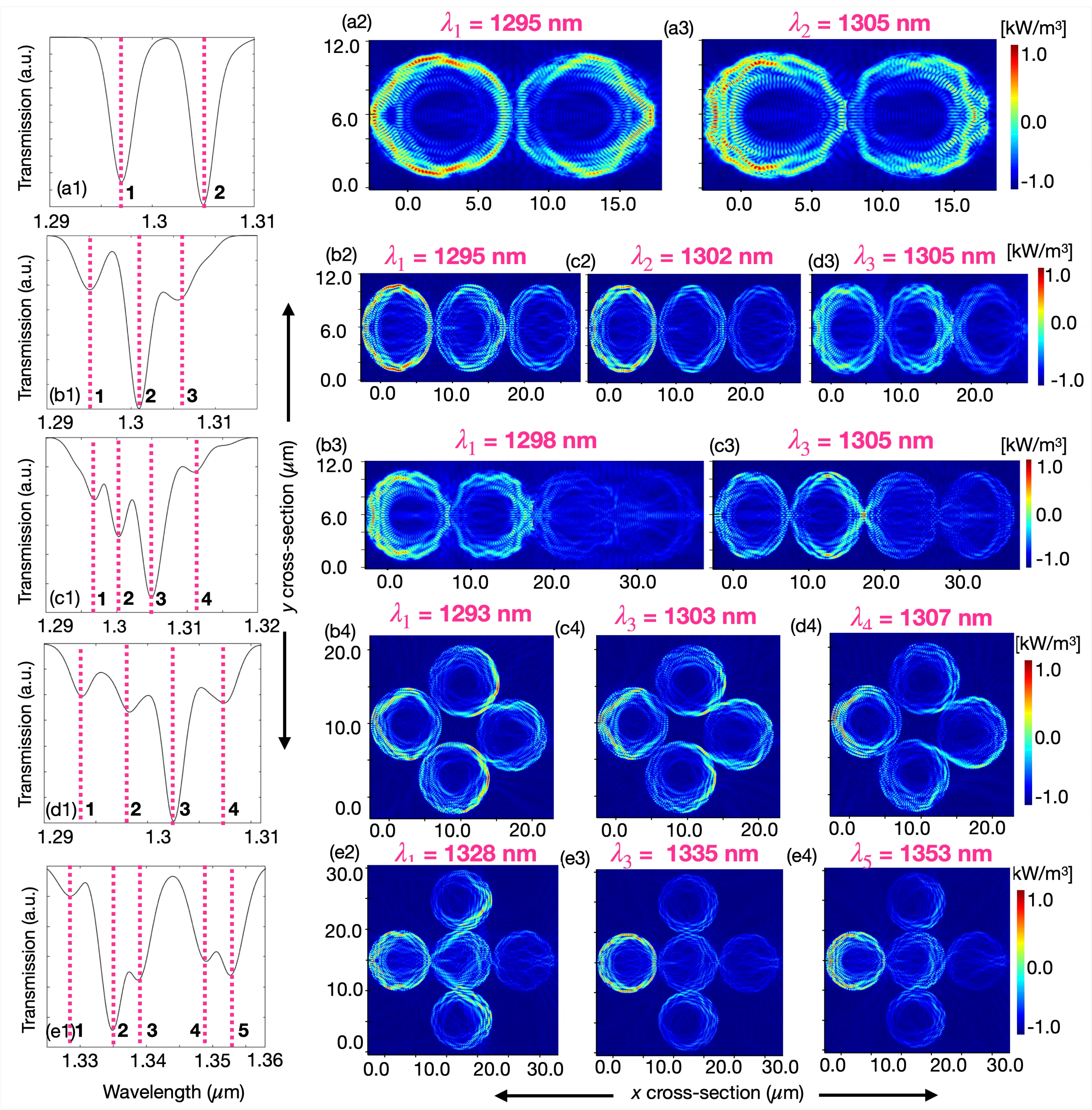}
     \caption{The electric-field distribution profiles of the photonic molecular systems calculated at the wavelengths corresponding to different supermodes. The following molecules are represented: linear chain formed by (a1) two circular, (b1) three circular and (c1) four circular resonators; (d1) quadrumer resonator, and (e1) five-cross circular resonator. Simulated EM ﬁeld maps at different supermode eigen wavelengths are presented for each molecule in (a2-a3), (b2-b4), (c2-c3), (d2-d4), and (e2-e4) respectively. }
     \label{fig:5}
 \end{figure*}

\subsubsection{Spatial distribution in $N$-high index Photonic molecular system:}
The electric field distributions corresponding to the dips in the transmission spectra provide deeper insight into the nature of supermodes in the photonic molecules, as shown in Figure~\ref{fig:5}. For the shortest wavelength in two-sphere system (Figure 5a2–5a3), the mode at ($\lambda$$_1$ = 1295 nm) shows a rather uniform field distribution across both cavities, thus reducing the field overlap at the junction of the cavities, corresponding to an antibonding state with weaker effective coupling. The field intensity is notably reduced at the contact points between adjacent cavities, indicating weakened direct evanescent coupling. In contrast at ($\lambda$$_2$ = 1305 nm), the field looks concentrated at the contact region, indicating an in-phase (bonding) supermode with strong inter-cavity coupling. 

As the number of resonators increases, the supermode structure becomes progressively richer due to coupling induced splitting. For the three-sphere chain (Figure~\ref{fig:5}(b1–b2)), three eigen states are observed, with the shortest wavelength mode showing a uniform field distribution (antibonding), while higher wavelength modes exhibit increasing asymmetry and localization at the junction, characteristic of bonding behavior. For larger systems (Figure~\ref{fig:5}(c–e)), the field patterns become more complex, with delocalized lower-order modes and asymmetric higher-order modes displaying nodes and intensity suppression. In compact clusters, multiple coupling pathways further distort the field distribution, leading to localized energy regions and the coexistence of bonding, antibonding, and weakly coupled (dark) states.

\begin{table} [htbp]
\caption{Quantitative comparison for $Q$-factors of two representative spectrally resolved supermodes obtained by Lorentzian fit, for 10 $\mu$m pristine photonic cavities.}
\small
\begin{tabular}{|p{1.1cm}|p{1.5cm}|p{2.5cm}|p{1.5cm}|p{1.5cm}|}
\hline
\textbf{$N$-Cavity} & \textbf{Mode} & \textbf{Representative Resonance($\lambda_\mathrm{res}$)} & \textbf{FWHM} & \textbf{$Q$-factor} \\ 
\hline
$N$ = 1 & - & 1297.40 nm &  1.736 nm & 0.747$\times10^3$ \\ 
\hline
$N$ = 2 & Mode 1 & 1297.00 nm & 1.336 nm & 0.970$\times10^3$ \\
 & Mode 2 & 1304.98 nm & 1.442 nm & 0.904$\times10^3$ \\
\hline
\multicolumn{5}{|p{8cm}|}{\textbf{Linear Topology}} \\ 
\hline
$N$ = 3 & Mode 1 & 1299.09 nm & 1.368 nm & 0.949$\times10^3$ \\ 
  & Mode 2 & 1304.92 nm & 1.478 nm & 0.882$\times10^3$ \\
\hline
$N$ = 4 & Mode 1 & 1296.10 nm & 1.088 nm & 1.190$\times10^3$ \\ 
 & Mode 2 & 1305.32 nm & 1.507 nm & 0.865$\times10^3$ \\
\hline
$N$ = 5 & Mode 1 & 1297.31 nm & 1.061 nm & 1.221$\times10^3$ \\ 
 &Mode 2 & 1303.55 nm & 1.190 nm & 1.095$\times10^3$  \\
\hline
$N$ = 6 & Mode 1 & 1303.31 nm & 0.980 nm & 1.329$\times10^3$ \\ 
 & Mode 2 & 1297.30 nm & 1.050 nm & 1.234$\times10^3$ \\
\hline
\multicolumn{5}{|p{8cm}|}{\textbf{Cluster Topology}} \\ 
\hline
$N$ = 3 & Mode 1 & 1298.02 nm & 1.090 nm & 1.190$\times10^3$ \\ 
 & Mode 2 & 1305.88 nm & 1.698 nm & 0.768$\times10^3$ \\
\hline
$N$ = 4 & Mode 1 & 1296.83 nm & 0.996 nm & 1.301$\times10^3$ \\
  & Mode 2 & 1304.49 nm & 1.454 nm & 0.897$\times10^3$ \\
\hline
$N$ = 5 & Mode 1 & 1299.28 nm & 1.513 nm & 0.858$\times10^3$ \\ 
  & Mode 2 & 1302.19 nm & 1.071 nm & 1.215$\times10^3$ \\
\hline
$N$ = 6 & Mode 1 & 1301.43 nm & 2.276 nm & 0.571$\times10^3$ \\
 & Mode 2 & 1307.00 nm & 0.970 nm & 1.34$\times10^3$ \\
\hline
\end{tabular}
\label{tab:I}
\end{table}

To assess whether the spectral restructuring induced by topology and dielectric-shell engineering is accompanied by significant resonance broadening, the $Q$-factor of representative, spectrally resolved hybridized modes were extracted from the simulated transmission spectra. \ref{tab:I} summarizes the resonance linewidths and corresponding $Q$-factors calculated. These values are obtained through Lorentzian fitting as described in Appendix B. The extracted $Q_\mathrm{spec}$ values vary with both molecular topology and the selected hybridized mode. For the linear configurations, the two representative modes yield $Q_\mathrm{spec}$ values of 0.970$\times$10$^3$ and 0.904$\times$10$^3$ for $N$=2, while the corresponding values for $N$=3, $N$=4, $N$=5, and $N$=6 remain within approximately 0.86$\times$10$^3$–1.33$\times$10$^3$. The compact configurations exhibit a broader mode-to-mode variation, with $Q_\mathrm{spec}$ ranging from 0.571$\times$10$^3$ to 1.34$\times$10$^3$. Thus, the linewidths are not governed solely by the number of resonators rather, they depend on the particular hybridized mode and the connectivity of the molecular topology. Nevertheless, the representative resonances remain spectrally well resolved with $Q_\mathrm{spec}$ values predominantly of order 10$^3$.


Thus, at shorter wavelengths ($\lambda_1$), the modes exhibit reduced field intensity at the contact regions and enhanced intensity along the outer periphery of the structure. A better preview of the field distribution profile, has been obtained from a smaller diameter (4 $\mu$m) cavity resonators, as shown in Figure~\ref{fig:A6} of Appendix C. The propagation of field confirms the existing theory on light coupling in resonators \cite{Li2015}, that inter-cavity coupling is not dominated solely by direct evanescent overlap at the contact points. This trapped light also involves propagation of scattered fields along the external boundary of the molecule. As a result, a larger fraction of the optical field resides in the surrounding medium, leading to a reduced effective refractive index and the observed shift of these modes toward shorter wavelengths, and hence notably higher $Q$-factor \cite{Li2017}. In contrast, modes appearing at longer wavelengths show strong field localization at the contact regions, with clear continuity of intensity across adjacent resonators. This confirms that coupling in these modes is dominated by phase-matched evanescent transfer between neighboring cavities, enabling efficient energy transport across the entire molecule, and hence bonding-type supermodes. These observations establish a direct correspondence between spectral position and coupling mechanism, where different wavelengths selectively excite supermodes governed by distinct spatial energy distributions.

\subsection{For coated spheres}
In this sub-section, we shall discuss the effect of coating a higher refractive index media (perovskite) upon the TiO$_2$ cavities. Within the coupling-detuning framework, the perovskite coating simultaneously increases the effective detuning ($\Delta$ = ($\lambda_i -\lambda_o$)/$\lambda_o$) and reduces the inter-cavity coupling parameter ($g$), enabling controlled tuning of hybridization without modifying geometry. As is the case before, the TiO$_2$ sphere diameters were fixed at 10 and 25 $\mu$m, while the perovskite shell thickness was varied (100 and 200 nm). As shown in Figure~\ref{fig:6}, the number of observable modes and their ordering remain consistent with the pristine systems, and the $Q$-factors ($\sim 10^3$) are largely preserved. However, the detuning values (shown in Figure~\ref{fig:A2} and Figure~\ref{fig:A3}) shift and compress, indicating that the coating primarily weakens inter-cavity coupling while modifying the local dielectric environment.

For linear chains, the uniform splitting is retained after coating, but with reduced inter-mode separation particularly for thicker (200 nm) perovskite layers, as shown in Figure~\ref{fig:A2} and Figure~\ref{fig:A3}, respectively. Despite supporting $N$ eigenmodes, only three dominant supermodes are resolved, as closely spaced modes near the band center merge due to near-degeneracy and finite linewidth, forming band-like spectral features. For 100 nm shells, the detuning range decreases, typically $\Delta$ $\sim$ -0.002 to +0.006, with nearly uniform spacing $\Delta_{12}$ $\approx$ $\Delta_{23}$ $\sim$ 0.003\text{–}0.004, consistent with preserved one-dimensional coupling behavior . Increasing the thickness to 200 nm further compresses the spectrum ($\Delta$ $\le$ 0.004), with reduced separations ($\sim$ 0.002\text{–}0.003) while maintaining relative ordering . This systematic reduction in spectral span directly reflects weakened evanescent overlap due to the intermediate shell.

In contrast, compact clusters show a more significant redistribution of detuning values upon coating. While the asymmetric splitting, which is a characteristic of non-linear configurations, is preserved, the central and outer modes shift differently depending on the coating thickness. This suggests that the perovskite layer introduces non-uniform modification of coupling pathways, particularly in multi-directional geometries where coupling is mediated through both contact regions and peripheral fields. As a result, the spectral fingerprints of clusters become slightly smoothened compared to the pristine case, although their topology dependent characteristics remain distinguishable. 

 \begin{figure*}
     \centering
     \includegraphics[scale=0.25]{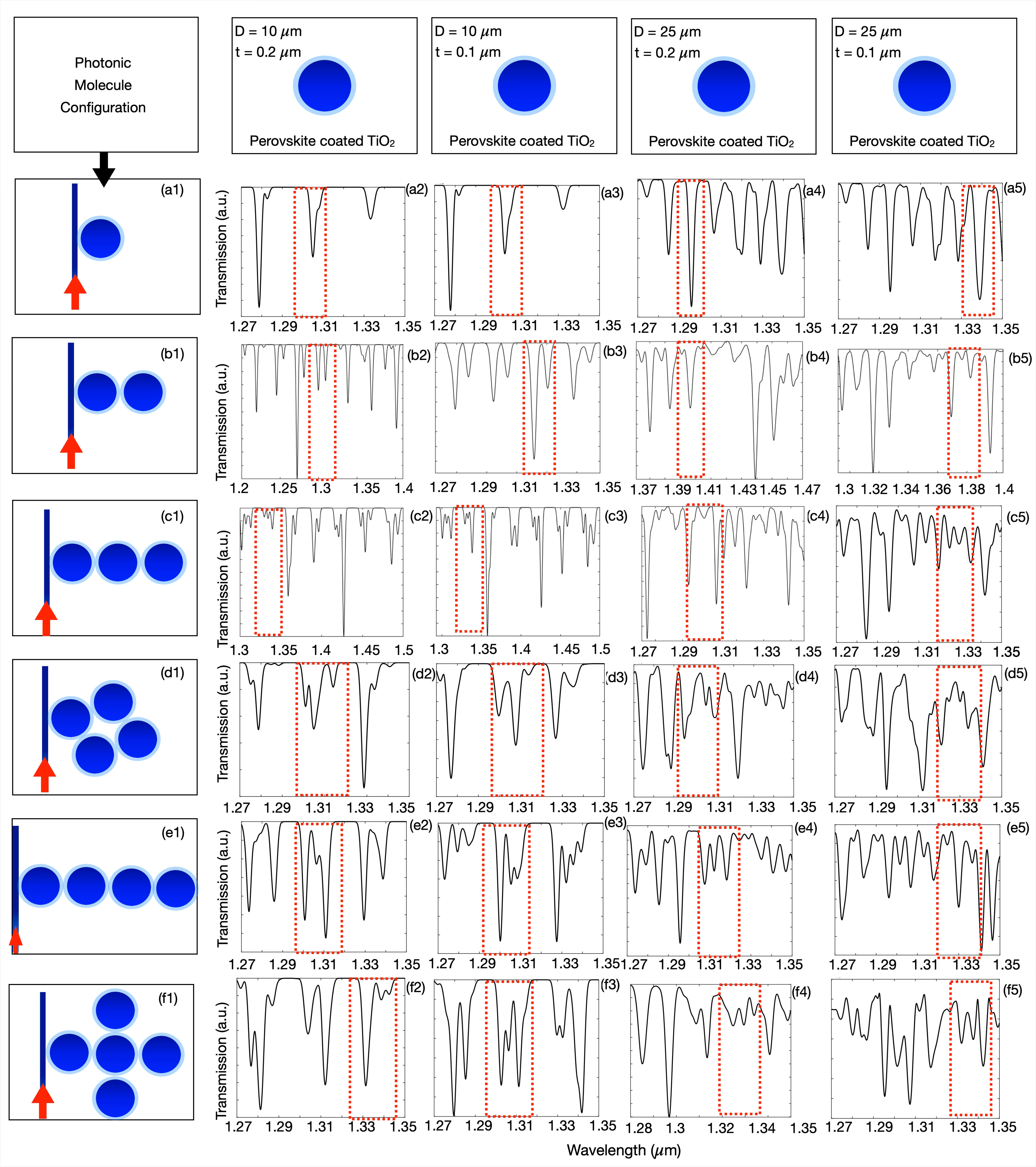}
     \caption{Simulated transmission spectra of hybridization in WGM, coupled optical supermodes are formed in perovskite coated TiO$_2$ core-shell photonic molecules. As a result, each uncoupled WGM resonance (a1-a5) forms a set of supermodes with splitting pattern for cavities with core diameter D=10 $\mu$m and shell thickness 200 nm (a2-f2) and 100 nm (a3-f3), and core diameter D=25 $\mu$m and shell thickness 200 nm (a4-f4) and 100 nm (a5-f5), with representative given molecular configurations illustrated in (a1-f1), respectively. The red dashed boxes state the spectra signatures in the molecular configurations including: (a1) single cavity, (b1) bi-cavity, linear chain of (c1) 3 circular resonators and (e1) 4 circular resonators, and planar (d1) quadrumer, and (f1) 5 cross resonator.}
     \label{fig:6}
 \end{figure*}

For thicker coatings (200 nm), the above-mentioned trends become more evident. In linear chains, the detuning range further compresses (within $\pm$ 0.004), and the separations decrease to ($\Delta \sim$ 0.002$\text{–}$0.003), while maintaining near-uniform spacing ($\Delta_{\max}/\Delta_{\min} \approx$ 1). In compact clusters, although asymmetric splitting persists, the spread between outer and central modes is reduced, and the detuning values are more closely clustered. This indicates that the differences between coupling pathways are suppressed with increasing coating thickness, driving the system toward a weaker coupling regime. Despite this compression, the persistence of identifiable mode patterns confirms that the photonic molecule structure remains intact, while the coupling strength can be tuned through the coating thickness.

\begin{table*} [htbp]
\caption{Quantitative comparison for $Q$-factors of two representative spectrally resolved supermodes obtained by Lorentzian fit, for 10 $\mu$m perovskite-coated photonic cavities.}
\small
\begin{tabular} 
{|p{2.5cm}|p{1.5cm}|p{2.5cm}|p{1.5cm}|p{1.5cm}|p{2.5cm}|p{1.5cm}|p{1.5cm}|}
\hline
\multicolumn{1}{|c|}{} &
\multicolumn{1}{|c|}{} &
\multicolumn{3}{c|}{\textbf{Perovskite Coating: 100 nm}} &
\multicolumn{3}{c|}{\textbf{Perovskite Coating: 200 nm}} \\
\hline
\textbf{Configuration ($N$)} &\textbf{Mode} &\textbf{Representative Resonance ($\lambda_\mathrm{res}$)} &\textbf{FWHM} & \textbf{$Q$-factor} &\textbf{Representative Resonance ($\lambda_\mathrm{res}$)} &\textbf{FWHM} & \textbf{$Q$-factor}  \\
\hline
$N$ = 1 & - & 1332.39 nm & 2.395 nm & 0.556$\times10^3$ & 1333.53 nm & 1.800 nm  &  0.741$\times10^3$ \\ 
\hline
$N$ = 2 & Mode 1 & 1328.53 nm & 1.595 nm &  0.833$\times10^3$ & 1329.31 nm & 0.941 nm & 1.411$\times10^3$ \\ 
& Mode 2 &  &  &  & 1333.47 nm & 1.467 nm & 0.91$\times10^3$   \\ 
\hline
\multicolumn{7}{|p{8cm}|}{\textbf{Linear Topology}} \\ 
\hline
$N$ = 3 & Mode 1 & 1360.61 nm & 1.573 nm & 0.864$\times10^3$ & 1329.31 nm & 1.073 nm & 1.237$\times10^3$ \\ 
& Mode 2 & 1369.07 nm & 2.034 nm & 0.673$\times10^3$ & 1334.19 nm & 0.985 nm & 1.353$\times10^3$ \\ 
\hline
$N$ = 4 & Mode 1 & 1327.54 nm & 1.618 nm & 0.82$\times10^3$ & 1329.42 nm & 1.397 nm  & 0.951$\times10^3$ \\ 
& Mode 2 & 1332.38 nm & 2.297 nm & 0.58$\times10^3$ & 1338.54 nm & 1.211 nm & 1.104$\times10^3$ \\
\hline
$N$ = 5 &  & & &  & &  \\ 
\hline
\multicolumn{7}{|p{8cm}|}{\textbf{Cluster Topology}} \\ 
\hline
$N$ = 3 & Mode 1 & 1330.81 nm & 1.429 nm & 0.934$\times10^3$ & 1331.98 nm & 1.276 nm & 1.043$\times10^3$ \\ 
& Mode 2 & 1337.21 nm & 1.348 nm & 0.991$\times10^3$ & 1335.01 nm & 1.817 nm & 0.734$\times10^3$ \\
\hline 
$N$ = 4 & Mode 1 & 1326.78 nm & 0.664 nm & 1.995$\times10^3$ & 1301.00 nm & 1.006 nm & 1.293$\times10^3$ \\ 
& Mode 2 &1335.75 nm & 1.808 nm & 0.739$\times10^3$ &1305.18 nm & 2.693 nm & 0.485$\times10^3$ \\
\hline
$N$ = 5 & Mode 1 & 1330.24 nm & 1.451 nm & 0.916$\times10^3$ & 1303.90 nm & 1.442 nm  & 0.904$\times10^3$ \\ 
& Mode 2 & 1341.19 nm & 2.616 nm & 0.512$\times10^3$ & 1312.18 nm & 1.902 nm & 0.69$\times10^3$ \\
\hline
\end{tabular}
\label{tab:II}
\end{table*}

\begin{table*} [htbp]
\caption{Quantitative comparison of the normalized detuning range of hybridized modes for pristine and perovskite-coated photonic molecules for 10 $\mu$m cavities. For each configuration, $\Delta_i$ shows the normalized detuning, and $\Delta_\mathrm{range}$ is the corresponding detuning range. The reference parent wavelength considered for pristine and perovskite-coated cavities with perovskite thickness 100 nm and 200 nm, are 1247.8 nm, 1302.79 nm and 1304.99, respectively. The relative change is subsequently calculated by comparing the detuning range of each coated configuration with that of the corresponding pristine photonic molecule having identical topology.}
\small
\begin{tabular} {|p{2.5cm}|p{1.5cm}|p{2.5cm}|p{1.5cm}|p{2cm}|p{1.5cm}|p{1.5cm}|p{2cm}|p{1.5cm}|}
\hline
\textbf{Configuration ($N$)} & \textbf{Pristine (Span)}  & \textbf{$\Delta_\mathrm{range}$(pristine)} & \textbf{100 nm (Span)} & \textbf{$\Delta_\mathrm{range}$(100 nm)} & \textbf{Relative Change} & \textbf{200 nm (Span)} & \textbf{$\Delta_\mathrm{range}$(200 nm)} & \textbf{Relative Change} \\ 
\hline
\multicolumn{9}{|p{16cm}|}{\textbf{Linear Topology}} \\ 
\hline
$N$ = 2 &  &  & 9.27 nm & 7.12e-3 &  & 9.28 nm & 7.11e-3 & \\ 
\hline
$N$ = 3 & 5.47 nm & 4.38e-3 & 11.17 nm & 8.57e-3 & -95.59\% & 7.43 nm & 5.69e-3 & -29.88\%\\ 
\hline
$N$ = 4 & 12.55 nm & 1.005e-2 & 8.50 nm & 6.52e-3 & 35.13\% & 10.19 nm & 7.808e-3 & 22.36\% \\ 
\hline
\multicolumn{9}{|p{16cm}|}{\textbf{Cluster Topology}} \\ 
\hline
$N$ = 3 & 16.12 nm & 1.291e-2 & 11.23 nm &  8.62e-3 & 33.28\% & 12.10 nm & 9.27e-3 & 28.23\% \\ 
\hline
$N$ = 4 & 14.52 nm & 1.163e-2 & 14.12 nm & 1.083e-2 & 6.86\% & 13.33 nm & 1.021e-2 & 12.22\% \\ 
\hline
$N$ = 5 & 28.50 nm & 2.284e-2 & 8.61 nm & 6.60e-3 & 71.06\% & 6.95 nm & 5.32e-3 & 76.78\% \\ 
\hline
\end{tabular}
\label{tab:III}
\end{table*}

\begin{figure*}
     \centering
     \includegraphics[scale=0.3]{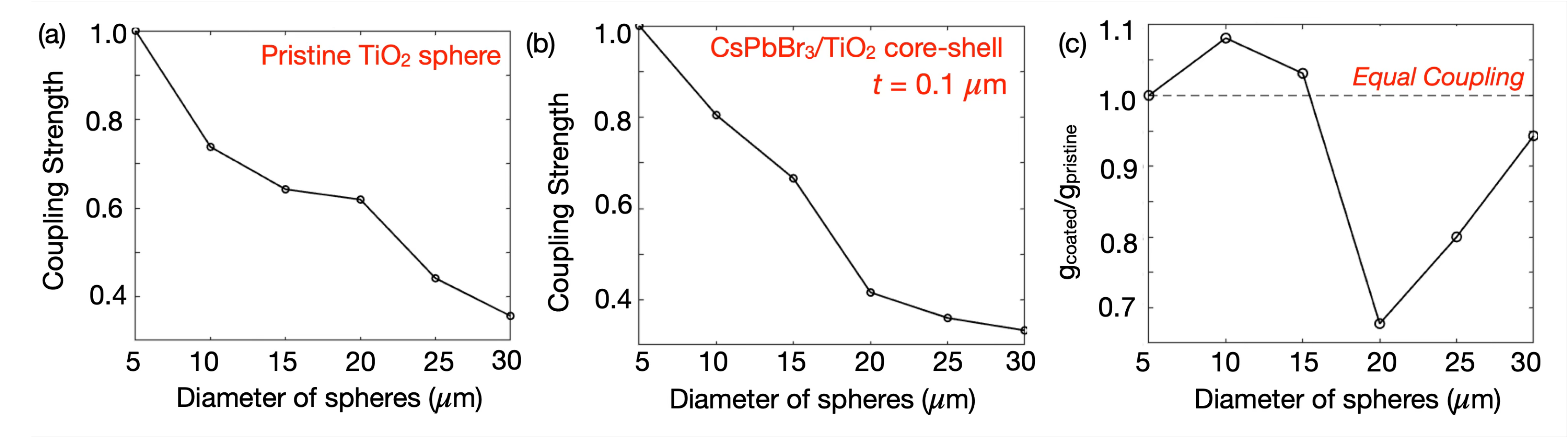}
     \caption{Diameter dependent coupling strength for (a) pristine and (b) perovskite-coated cavities with perovskite thickness $t$ = 100 $nm$, and (c) Effective coupling of both the configurations in bi-cavity circular resonator system.}
     \label{fig:7}
\end{figure*}

Compared to pristine TiO$_2$ photonic molecules, the perovskite-coated systems exhibit a reduction in coupling strength, reflected in the compressed detuning range and decreased inter-mode separations. In the pristine case, the spectral span typically extends over ($\Delta \sim \pm$ 0.01 $\text{–}0.02$), with larger mode splitting and stronger distinction between adjacent supermodes. In contrast, for coated systems (100 nm and 200 nm), the detuning range reduces to ($\Delta \sim \pm$ 0.004$\text{–}0.01$), and the adjacent mode separations decrease to ($\Delta \sim$ 0.002$\text{–}0.004$), indicating weakened hybridization between resonators. This effect is more pronounced for thicker coatings, confirming that the perovskite layer reduces evanescent overlap and effectively decouples the cavities. 

The effect of dielectric-shell engineering on the spectral linewidths was further examined by extracting the $Q$-factors of representative resonances for the 100 and 200 nm perovskite-coated photonic molecules, as summarized in Table \ref{tab:II}. For the 100 nm coating, the extracted $Q$-factor values of the coupled configurations span approximately 0.58$\times10^3$–0.99$\times10^3$, except for the isolated cavity with $Q$=0.556$\times10^3$. With a 200 nm shell, the corresponding values extend from approximately 0.49$\times10^3$ to 1.41$\times10^3$. The variation between the two fitted modes within a given topology demonstrates that the shell does not simply produce a uniform change in resonance linewidth, rather, its effect depends on the hybridized state and molecular connectivity. Importantly, the resonances remain spectrally resolved across the investigated configurations.


To quantitatively compare the spectral redistribution induced by dielectric-shell incorporation, as shown in Table \ref{tab:III} the normalized detuning of each hybridized mode is defined as: $\Delta_i$ = ($\lambda_i -\lambda_o$)/$\lambda_o$, where $\lambda_o$ is the parent WGM resonance and $\lambda_i$ is the wavelength of the corresponding hybridized mode. For a given molecular configuration, the spectral extent of the hybridized modes is quantified by the corresponding detuning range, $\Delta_\mathrm{range}$ = $\Delta_\mathrm{max}$ - $\Delta_\mathrm{min}$. The relative spectral change (= 1 - $\frac{\Delta_\mathrm{range}\mathrm{(coated)}}{\Delta_\mathrm{range}\mathrm{(pristine)}} \times 100 \%$) induced by the dielectric shell is evaluated with respect to the corresponding pristine photonic molecule with the same molecular topology and resonator arrangement.

To further elucidate the spatial characteristics of hybridized modes in coated systems, additional field distribution profiles are provided in the Appendix B (Figure~\ref{fig:A5} - Figure~\ref{fig:A7}). For larger cavity diameters (10 and 25 $\mu m$), the strong confinement of WGMs near the resonator surface results in closely spaced and less distinguishable field patterns across hybridized modes, making clear visualization of modal evolution challenging in full-wave simulations. Therefore for better visualization and clarity, the representative field distributions are presented for smaller diameter systems (D = 4 $\mu m$), where the extended evanescent fields enhance inter-cavity interaction and allow clearer resolution of bonding, antibonding, and partially localized supermodes. Despite the difference in absolute cavity size, the qualitative features of hybridization, including mode symmetry, field localization, and topology-dependent distribution remain consistent with those observed in larger systems. 

\section{Coupling parameter variation for supermode analysis}
In addition to the case of 4 $\mu$m diameter cavities, the coupling behavior in pristine as well core-shell bisphere system has been studied for various diameters in the range 5–30 $\mu m$ are studied, and the results are summarized in Figure~\ref{fig:7}. For smaller diameters (5–15 $\mu m$), the WGMs are weakly confined within the cavities. This results in extended evanescent fields that enable efficient inter-cavity coupling \cite{Svetlana2006}. This is evident from the Figure~\ref{fig:7}(a and b), where the coupling parameter `$g$', decreases on increasing the diameter of the TiO$_2$ cavities from 5 $\mu m$ to 30 $\mu m$. However, there is a distinctive difference in the coupling behavior between two pristine cavities and two core-shell cavities with perovskite thickness $t =$ 100 $nm$, as seen in Figure~\ref{fig:7}(c). The non-monotonic variation of the coupling ratio $g_{\text{coated}}/g_{\text{pristine}}$ with cavity diameter, reflects the interplay between mode confinement and evanescent coupling. 

For smaller diameters (5–15 $\mu m$) regime, the perovskite coating modifies the refractive index environment and enhances the effective field overlap, leading to a slight increase in coupling strength compared to the pristine case. However, as the core diameter increases ($\approx$ 20 $\mu m$), the modes become strongly confined near the resonator surface, significantly reducing the evanescent field penetration. This shows that the introduction of the perovskite layer then acts as a dielectric spacer between the high-index cores, suppressing the overlap and leading to a sharp reduction in coupling strength. At larger diameters, a partial recovery is observed, likely due to increased contribution from peripheral coupling pathways. The similar trend in behavior has also been observed for a perovskite thickness of $t =$ 200 $nm$, as shown in Figure~\ref{fig:A4} of Appendix B. These results indicate that the effect of coating on coupling is strongly diameter dependent, with an optimal regime where material modification enhances interaction, followed by a transition to a barrier dominated regime where coupling is suppressed.

Thus, the pristine configuration is advantageous for applications requiring strong coupling, large mode splitting, and distinct spectral signatures, such as sensing and topology identification. In contrast, the coated system offers reduced coupling with improved control over field distribution and interaction with the surrounding medium, making it more suitable for applications involving light–matter interaction, absorption enhancement, or hybrid photonic–optoelectronic functionality. The choice between the two therefore depends on whether strong photonic interaction or controlled material interaction is the primary objective.
 
\section{Resonating modal analysis in  photonic molecule system}
\subsection{Single sphere resonating sphere}
Whispering gallery modes in dielectric spheres act as an adequate model, for studying the eigenmodes of optical cavities. For this we need to solve Maxwell's equations in dielectric media, which reduces to, Helmholtz equation 
\begin{align}\label{Eq:Helmholtz Eq}
    \nabla^2\mathbf{E}+k^2\mathbf{E}=0
\end{align}
satisfying $\nabla \cdot \mathbf{E} =0$. Instead of directly solving Eq.\ref{Eq:Helmholtz Eq}, we construct the field $\mathbf{E}$ out of a scalar generating functional $\mathbb{E}$ satisfying :
\begin{align}\label{Eq:Scalar_Helmholtz Eq}
    \nabla^2\mathbb{E}+k^2\mathbb{E} = 0
\end{align}

The generating functional $\mathbb{E}$ can be obtained via separation of variables, in the following form:
\begin{align} \label{Eq:Solutions_helhmoltz}
\mathbb{E}_{l,m}(r,\theta,\phi) = Z_l(kr)Y_{l}^m(\theta, \phi),
\end{align}
where, $Z_l(kr)$ follows the Bessel's equation in the radial direction and $Y_{l}^m(\theta, \phi)$ represents the spherical harmonics. In order to construct a divergence-free vector field out of this generating functional, we define:
\begin{align}
   & \mathbf{M}_{l,m} \equiv \mathbf{\nabla} \times (\mathbb{E}_{l,m}(r, \theta, \phi)\,\,\mathbf{r})  \label{Eq:M_equation}\\
   & \mathbf{N}_{l,m} \equiv \frac{1}{k}\left(\mathbf{\nabla} \times \mathbf{M}_{l,m}\right)
\end{align}
and the solution to Eq.~\ref{Eq:Helmholtz Eq} can be written as
\begin{align}\label{Eq:sol vec helmholtz}
    \mathbf{E}(\mathbf{r}) =\sum_{l,m} \left(A_{l,m} \mathbf{M}_{l,m} + B_{l,m} \mathbf{N}_{l,m}\right)
\end{align}
where $\mathbf{M}_{l,m}$ and $\mathbf{N}_{l,m}$ are tangential vector spherical harmonics (VSH) basis representing transverse electric (TE) and transverse magnetic (TM) fields, respectively. The explicit form of them are
\begin{align}
    \mathbf{M}_{l,m} & = Z_l(kr)\left[\frac{1}{\sqrt{l(l+1)}}\left(\mathbf{r} \times \nabla Y_l^m(\theta,\phi)\right)\right] \label{Eq:explicit form of M }\\
    \mathbf{N}_{l,m} & = \frac{l(l+1)}{kr} Z_l(kr)Y_l^m \hat{\mathbf{r}} + & \nonumber\\
   & \quad \quad \quad \quad \frac{1}{kr} \frac{d}{dr}\left(r Z_l(kr)\right)(r\nabla Y_l^m(\theta,\phi) )\label{Eq:explicit form of N }
\end{align}
Eq.~\ref{Eq:sol vec helmholtz} can be written in a compact form as:
\begin{align}
\mathbf{E}(r,\theta,\phi) = \sum_{\beta,l,m} C^{\beta}_{l,m} E^{\beta}_{l,m}(Z_l (kr), \theta, \phi),
\end{align}
where $C^{\beta=1}_{l,m} = A_{l,m}$, $C^{\beta=2}_{l,m} = B_{l,m}$, $E^{\beta=1}_{l,m} = \mathbf{M}_{l,m}$ and $E^{\beta=2}_{l,m} = \mathbf{N}_{l,m}$ and explained in Appendix D. As we previously mentioned, the radial equation for ($Z_l (kr)$) follows Bessel's equation; therefore, the inside solutions are written in terms of the Bessel function ($j_l(kr)$), and the outside solutions are written in terms of the Hankel function of the first kind ($h_l^{(1)}(kr)$). Hence, 
\begin{subequations}\label{Eq:field inside and outside}
\begin{align}
&\mathbf{E}_\mathrm{int}(r,\theta,\phi) = \sum_{\beta, l,m}a^{\beta}_{l,m} E^{\beta}_{l,m}(j_l (k_ir), \theta, \phi),
\label{Eq:inside filed} \\
&\mathbf{E}_\mathrm{sca}(r,\theta,\phi) = \sum_{\beta,l,m} b^{\beta}_{l,m} E^{\beta}_{l,m}(h_l^{(1)}(k_or), \theta, \phi)
\label{Eq:outside field}
\end{align}
\end{subequations}
Here, $k_o$ and $k_i$(=$\sqrt{\epsilon_i}k_o$) are the wave numbers in the vacuum and within the sphere $i$, respectively. The WGM are obtained by implementing the resonance condition at the boundary and are characterized in terms of mode numbers ($l,m$), where $l$ denotes the polar mode number, $m$ denotes the azimuthal mode number. The explicit expressions for coefficients $a^{\beta}_{l,m}$ and $b^{\beta}_{l,m}$ are expressed in Appendix E. 

Thus, from the source free Maxwell eigenvalue equations with outgoing wave boundary conditions, the optical resonances of the microcavity are obtained. This yields a complex eigen frequency ($ \omega_i=ck_i$), where the real part determines the oscillation frequency of the mode, and the imaginary part  is responsible for the radiative and absorptive losses in the cavity \cite{Franchimon2013}. From this eigenvalue, the measurable quantities are defined. This includes resonance wavelength ($\lambda_r = 2\pi c/{\rm{Re}}(\omega_i)$), and the quality factor ($Q=|{\rm{Re}}(\omega_i)/2{\rm{Im}}(\omega_i)|$), describing the ratio of energy stored to energy lost per cycle \cite{Atvars:21}. The corresponding spectral linewidth ($\Delta\lambda$ = $\lambda_r$/$Q$) represents the full width at half maximum of the resonance peak and is directly related to the photon lifetime. For a fixed polar mode $l$, each solution as in Eq.~\ref{Eq:inside filed} and \ref{Eq:outside field} corresponds to an azimuthally circulating electromagnetic wave confined by total internal reflection, which slowly leaks energy ($E(t)$) outward and decays exponentially in time as $E(t)\sim e^{i\omega_i t}$. Due to the azimuthal symmetry of the cavity, all WGMs possess $2l+1$ fold degeneracy sharing the same complex frequency. However, this degeneracy is lifted in the presence of external perturbations. Therefore, it presents a viable framework for interpreting the modes extracted from numerical simulations under the influence of perturbation parameters such as geometric asymmetry, material absorption and coupling to adjacent structures, in modifying the resonance properties of complex device configurations.

\subsection{Multi-sphere resonating spheres}
As has been discussed in section ~\ref{sec:numerical simulations}, when two spheres are brought into proximity, the evanescent modes of the second sphere act as a perturbation to the WGM of the first sphere and thereby lift the degeneracy. Therefore, we analyze the modified WGMs of the first sphere, in the presence of the second sphere by employing the single mode tight-binding (SMTB) formulation of Miyazaki and Jimba \cite{Miyazaki2000}, which projects the full Maxwell boundary integral equation onto the single sphere resonance basis. This produces a reduced non-Hermitian eigenvalue equation whose solutions correspond to the complex resonance frequencies of the bisphere photonic molecule. The secular eigenvalue equation captures both bonding and antibonding mode splitting, providing a quantitative description on the behavior of photonic molecules with the originally degenerate WGMs evolving into distinct supermodes when the spheres are brought into proximity. 

\subsubsection{Tight-binding formulation}
We consider a linearly polarized plane wave to be incident on a system of non-magnetic bi-spheres separated by distance `$d$'. The spheres with radius $a_i$ and permittivity $\epsilon_i$ are centered at $R_i$ with $i \in \lbrace1,2\rbrace$.
In the bisphere formulation, we retain the same internal and scattered field representations as in the single sphere (Mie) resonances. Following Eq.~\ref{Eq:inside filed} and \ref{Eq:outside field} the incident, internal and scattered fields can thus be expressed as:

\begin{subequations} \label{Eq:bisphere_field}
\begin{align}
&\mathbf{E}^{(i)}_{\mathrm{inc}}(r_i) = \sum_{\nu} c^{(i)}_{\nu} E^{(i)}_{\nu}(j_l(k_or_i), \theta, \phi),\label{Eq:incident bi sphere field}\\
&\mathbf{E}^{(i)}_{\mathrm{int}}(r_i) = \sum_{\nu} a^{(i)}_{\nu} E^{(i)}_{\nu}(j_l(k_ir_i), \theta, \phi),\label{Eq:inside bi sphere field} \\
&\mathbf{E}^{(i)}_{\mathrm{sca}}(r_i) = \sum_{\nu} b^{(i)}_{\nu} E^{(i)}_{\nu}(h_l^{(1)}(k_or_i), \theta, \phi).\label{Eq:outside bi sphere field} 
\end{align}
\end{subequations}
For notational convenience, we relabeled the indices $(l,m,\beta)$ as $\nu$ and introduced shifted coordinates $r_i=r-R_i$. 
The boundary conditions at the surface of sphere $i$ are
\begin{align} \label{Eq:boundary_conditions}
\hat{\mathbf{r}} \times \mathbf{E}_{\mathrm{int}}^{(i)}(r_i)= \hat{\mathbf{r}} \times \mathbf{E}^{(i)}_{\mathrm{inc}}(r_i)+ \hat{\mathbf{r}} \times \mathbf{E}^{(j)}_{\mathrm{sca}}(r_i)
\end{align}
where, $\mathbf{E}^{(j)}_{\mathrm{sca}}(r_i)$ represents the field scattered by sphere `$j$' evaluated at $r_i$. As we are applying the boundary conditions Eq.~\ref{Eq:boundary_conditions} at the surface of sphere $i$, therefore, we need to expand  $\mathbf{E}^{(j)}_{\mathrm{sca}}(r_i)$  in the VSH basis of the sphere $i$ which can be written as 
\begin{align}
\mathbf{E}_{\mathrm{sca}}^{(j)}(r_i)= \sum_{\nu}\left(\sum_{\nu'}\mathcal{T}^{(j\rightarrow i)}_{\nu\nu'}(\omega,d)b_{\nu'}^{(j)} \right)E^{(i)}_{\nu}(h_{l}^{(1)}(k_or_i)).
\end{align}
Here, $b_{\nu'}^{(j)}$ is the amplitude of the mode $\nu'$ scattered from sphere $j$ and
$\mathcal{T}^{(j\rightarrow i)}_{\nu\nu'}$ represents how much mode $\nu'$ from sphere $j$ becomes mode $\nu$ when measured from the sphere $i$. $\mathcal{T}^{(j\rightarrow i)}_{\nu\nu'}$ are called translation coefficients and are obtained by the direct evaluation of Wigner 3-$j$ symbols \cite{CRUZAN1962} which are summarized in Eq.~\ref{Eq:Wigner_3j} of Appendix G. The total incident field coefficients of sphere $i$ then becomes the sum of externally incident field and the scattered field from sphere-$j$. Therefore, the coefficient $c^{(i)}_{\nu}$ in Eq.~\ref{Eq:incident bi sphere field} can be expressed as,
\begin{align}\label{Eq:cnu}
c^{(i)}_{\nu}=f^{(i)}_{\nu}+\sum_{\nu'}\mathcal{T}^{(j\rightarrow i)}_{\nu\nu'}(\omega,d)b^{(j)}_{\nu'}
\end{align} 
where, $f^{(i)}_{\nu}$ is the coefficient of externally incident field on sphere $i$. 

We introduce the response operators ($R^{(i)}_{\nu\nu'}$ and $S^{(i)}_{\nu\nu'}$) for the sphere-$i$ as,

\begin{align}
&a^{(i)}_{\nu}=\sum_{\nu'} R^{(i)}_{\nu\nu'}(\omega)c^{(i)}_{\nu'}, \label{Eq:anu}\\
&b^{(i)}_{\nu}=\sum_{\nu'} S^{(i)}_{\nu\nu'}(\omega)a^{(i)}_{\nu'}\label{Eq:bnu}.
\end{align} 
where $R^{(i)}$ is the internal field response operator and $S^{(i)}$ is the Mie scattering operator. The expression for the scattering operator are discussed in Eq. \ref{eq:Mie scattering in prsitine spheres} and \ref{eq:Mie scattering in core-shell spheres}. Substituting Eq.~\ref{Eq:anu} and Eq.~\ref{Eq:bnu} in Eq.~\ref{Eq:cnu} we obtain the scattering coefficient for both the spheres as: 
\begin{align}\label{Eq:bnu final}
b^{(i)}_{\nu}=\sum_{\nu'} S^{(i)}_{\nu\nu'}(\omega)\left[f^{(i)}_{\nu'}+\sum_{\mu}\mathcal{T}^{(j\rightarrow i)}_{\nu'\mu}(\omega,d)b^{(j)}_{\mu}\right]
\end{align}
and the reshuffling in terms results,
\begin{align}\label{Eq:reshuffle bnu final}
\sum_{\mu}&\delta_{\nu\mu}b_{\mu}^{(i)}-\sum_{\mu}\left(\sum_{\nu'}S^{(i)}_{\nu\nu'}(\omega)\mathcal{T}^{(j\rightarrow i)}_{\mu\nu'}(\omega,d)\right) b^{(j)}_{\mu} \nonumber\\
& \qquad\qquad\qquad\qquad \qquad = \sum_{\mu} S^{(i)}_{\nu\mu}(\omega)f^{(i)}_{\mu} 
\end{align}
where $\delta_{\nu\mu}$ is the Kronecker delta and represents the response of a non-interacting single sphere, whereas the product $S^{(i)}\mathcal{T}^{(j\rightarrow i)}$ refers to the modification in the field response of the first sphere, induced by the presence of the second sphere. This provides the self-consistent condition that governs the evanescent field interaction between the two spheres. The expressions for Mie coefficients for pristine and core-shell spheres have been discussed in Appendix E and Appendix F, respectively. 

For computational convenience, we rewrite the variables and equations in a matrix form as follows:
\begin{subequations}\label{Eq:Matrix defn}
    \begin{align}
&\mathbf{b}_\nu=\begin{pmatrix}
b^{(1)}_{\nu}\\
0\\
b^{(2)}_{\nu}\\
0
\end{pmatrix}, \quad \mathbf{f}_\nu=\begin{pmatrix}
f^{(1)}_{\nu}\\
0\\
f^{(2)}_{\nu}\\
0
\end{pmatrix}, \label{Eq: bnu fnu }\\
&\mathbf{I}_{\nu}=\begin{pmatrix}
    \delta_{\mu\nu} & 0\\
    0 & \delta_{\mu\nu}
    \end{pmatrix}, \quad \mathbf{V}_{\nu}(\omega,d)= \sum_{\mu}\begin{pmatrix}\mathbf{0}_{2\times 2} & \mathcal{V}^{12}_{\mu\nu}, \\
\mathcal{V}^{21}_{\mu\nu} & \mathbf{0}_{2\times 2}
\end{pmatrix}, \label{Eq: Inu Vnu }\\
&\mathbf{S}_{\nu}(\omega)=\sum_{\mu}\begin{pmatrix} S^{(1)}_{\mu\nu}(\omega)\,\, \mathbb{I}_{2\times 2} & \mathbf{0}_{2\times 2} \\
\mathbf{0}_{2\times 2} & S^{(2)}_{\mu\nu}(\omega)\,\, \mathbb{I}_{2\times 2}
\end{pmatrix}.
\label{Eq:Mie scattering matrix}
\end{align}
\end{subequations}

where, 
\begin{subequations}\label{Eq:Definition mathcal V}
\begin{align}
    &\mathcal{V}^{12}_{\mu\nu} = \sum_{\nu'}S^{(1)}_{\nu\nu'}(\omega)\mathcal{T}_{\nu'\mu}^{(2\rightarrow 1)} (\omega,d)\,\, \mathbb{I}_{2\times 2},\label{Eq:Definition mathcal V1}\\
    &\mathcal{V}^{21}_{\mu\nu} = \sum_{\nu'}S^{(2)}_{\nu\nu'}(\omega)\mathcal{T}_{\nu'\mu}^{(1\rightarrow 2)} (\omega,d)\,\, \mathbb{I}_{2\times 2},\label{Eq:Definition mathcal V2}
\end{align}
\end{subequations}
Therefore, the matrix form of the Eq.~\ref{Eq:reshuffle bnu final} takes the following form
\begin{align}\label{Eq:Matrix form of chracteristic eq} 
    \big[\mathbf{I}_\nu - \mathbf{V}_{\nu}(\omega,d)\big] \mathbf{b}_{\nu} = \mathbf{S}_{\nu}(\omega) \mathbf{f}_{\nu}.
\end{align}
As $\mathbf{f}_{\nu}$ contains the projections of the external field onto the VSH basis, it vanishes for naturally occurring resonances. Hence, the characteristic equation takes the form:
\begin{align}
\det\big[\mathbf{I}_\nu - \mathbf{V}_{\nu}(\omega,d)\big] = 0.
\label{Eq:secular_full}
\end{align}
which on solving for the natural WGM resonances in identical spheres under single-mode approximation, simplifies to:
\begin{align}
1 - S_{\nu_0}^2(\omega)\mathcal{T}_{\nu_0}^2(\omega,d) = 0,
\label{Eq:final_secular}
\end{align}
where, $\nu_0$ represents the dominant resonant multipole for isolate WGM resonances. The detailed derivation has been presented in Appendix H. The solution to this secular equation~\ref{Eq:final_secular}, yields the split mode eigen frequencies, given by:
\begin{align}
\omega_{\pm} = \omega_0 \pm g(d).
\end{align}
where $\omega_0$ is the isolated single-sphere frequency,  $\omega_+$ represents the bonding mode and $\omega_-$ being the antibonding mode. The term $g(d)$ is the coupling parameter which is introduced as the effective strength, obtained from the secular equation near the isolated resonance, explicitly defined from $S_{\nu_0} = \frac{N_{\nu_0}(\omega_0)}{D_{\nu_0}(\omega_0)}$, given as:
\begin{align} \label{Eq:coupling parameter}
g(d) = \frac{N_{\nu_0}(\omega_0)}{D'_{\nu_0}(\omega_0)}\mathcal{T}_{\nu_0}(\omega_0,d)
\end{align}
The form of $N_{\nu_0}$ and $D_{\nu_0}$ is given in \ref{Eq:numerator/denominator} Appendix H. The parameter $g$ shows that the coupling is governed by the product of intrinsic scattering response operator and the spatial overlap between the modes, derived from $\mathbf{V}_{\nu}(\omega,d)$ in Eq.~\ref{Eq:Matrix form of chracteristic eq}.

\subsubsection{Tight-binding model vs numerical simulations}
The emergence of two nondegenerate solutions from a previously degenerate one is the mode splitting phenomenon characteristic of coupled resonators. Figure~\ref{fig:8} shows the spectral output of the tight-binding model against the full wave numerical simulations for both single as well as coupled TiO$_2$ microspheres systems. For a 10 $\mu m$ dia single sphere, the spectra obtained from the two methods exhibit an excellent correspondence (Figure~\ref{fig:8}(a)-(b)). Both approaches reproduce the resonance positions (resonances at (38-41,0) and coupling-induced splitting, with the tight-binding resonances appearing at the same size parameters range as those simulated in the FDTD transmission. This agreement confirms that the tight-binding model captures the eigenvalue structure of WGMs and provides a reliable analytical basis for describing modal evolution in composite resonator systems.

The FDTD and SMTB descriptions employ different geometrical representations of the resonator system. The FDTD calculations use a two-dimensional cylindrical representation of the spherical resonators, whereas the analytical SMTB treatment is formulated using the three-dimensional spherical VSH description. The comparison is therefore made at the level of the common WGM observables, namely the resonance positions and coupling-induced spectral splitting, rather than as a strict equivalence of the complete electromagnetic solutions. As shown in Figure~\ref{fig:8}, the SMTB treatment reproduces the resonance positions and the bonding–antibonding splitting obtained from the FDTD calculations, thereby validating the coupling description used in the subsequent analysis. Furthermore, the coupling strength can be directly related to the coupling-induced splitting of the two-resonator system, $\Delta \omega$=2$g$($d$), allowing the SMTB coupling coefficient to be compared with the splitting independently obtained from the FDTD spectra. The present validation is therefore concerned primarily with the resonance frequencies and coupling-induced spectral splitting.

The tight-binding spectra shown in this work were obtained by numerically evaluating the single sphere Mie resonance operators ($S^{(i)}(\omega)$) and the VSH translation operators ($\mathcal{T}^{(j\to i)}(\omega,d)$), followed by construction of the non-Hermitian bisphere Hamiltonian as discussed above. The roots of the secular determinant have been computed over the frequency (or equivalently size parameter) range, yielding the two supermode eigenfrequencies ($\omega_{\pm}$). These frequencies have been mapped to effective wavelength positions and compared directly with the FDTD extracted mode peak positions. The Mie operators have been computed using spherical Bessel and Hankel functions with argument ($x = 2\pi a/\lambda$), where $a$ is the sphere radius. The translation operators has been evaluated using the VSH addition theorem, including the full angular momentum coupling through Wigner $3-j$ coefficients, and the separation between the spheres has been set to the touching condition ($d = 2a$).

 \begin{figure}
     \centering
     \includegraphics[scale=0.18]{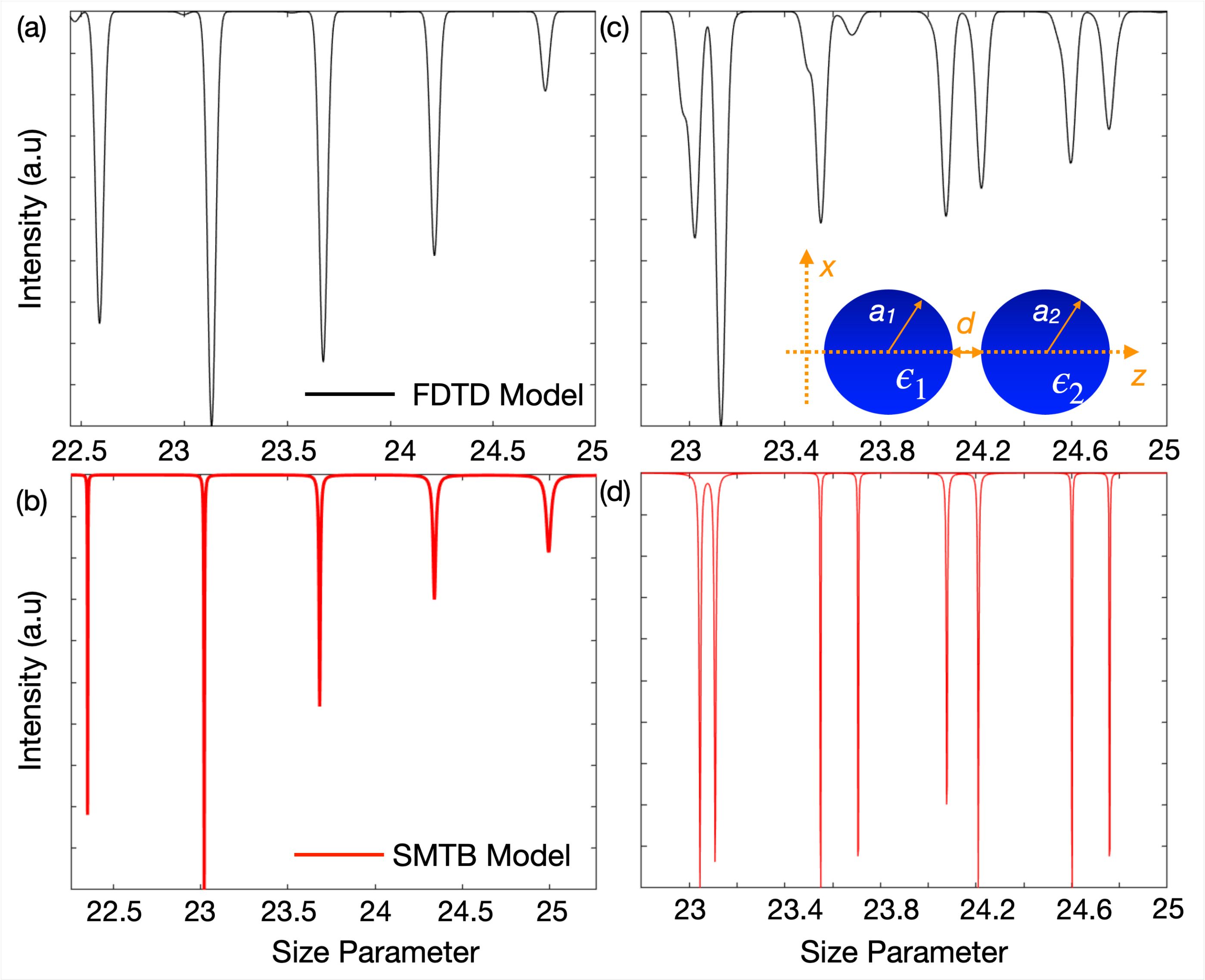}
     \caption{Comparison of intensity vs. the size parameter (= $2\pi a/\lambda$) curves for 10 $\mu$m diameter TiO$_2$ spheres through FDTD simulations and single mode tight-binding model (SMTB) model. Transmission spectra obtained from FDTD simulations for (a) single spheres and (c) double spheres systems; (inset) Double sphere aligned along the z-axis, centred at $R_i$, separated by distance $d$. Each sphere has radius $a_i$ and dielectric constant $\epsilon_i$. Intensity spectra obtained from the SMTB model for (b) single spheres and (d) double spheres. Although the SMTB formulation yields complex eigenfrequencies and consequently contains information on resonance decay, a separate decomposition of radiative, absorptive, and external-coupling loss channels is not performed.}
     \label{fig:8}
 \end{figure}

As discussed in the above section, when two resonating cavities are brought in close contact, each originally isolated WGMs split into a pair of hybridized resonances. The FDTD spectrum in Figure~\ref{fig:8}(c) shows a spectra of closely spaced mode doublet dips. These doublets correspond to the symmetric (bonding) and antisymmetric (antibonding) supermodes of the photonic molecule, formed through coherent hybridization of the evanescent WGM fields localized near the contact region of the bispheres. The tight-binding model derived spectra in Figure~\ref{fig:8}(d) reproduces the similar trend in both mode ordering and splitting magnitude. For each parent WGM, the tight-binding formulation returns two distinct eigenfrequencies ($\omega_{\pm}$), and these align with the FDTD doublets across the entire mode range. Furthermore, for larger photonic molecules, the same pairwise coupling framework is generalized to an $N \times N$ Hamiltonian (See Figure \ref{fig:A9} and Section J of Appendix), with nonzero off-diagonal elements assigned only to nearest-neighbor resonators. The resulting eigenvalues represent the collective supermodes of the molecule. The SMTB treatment is employed here to identify the topology-dependent hybridization mechanism rather than to reproduce the full electromagnetic spectrum. In particular, the present Hermitian nearest-neighbor formulation does not explicitly incorporate the radiative and absorptive loss channels of the high-index coated resonators, which are captured directly by the FDTD calculations.

  \begin{figure}
     \centering
     \includegraphics[scale=0.24]{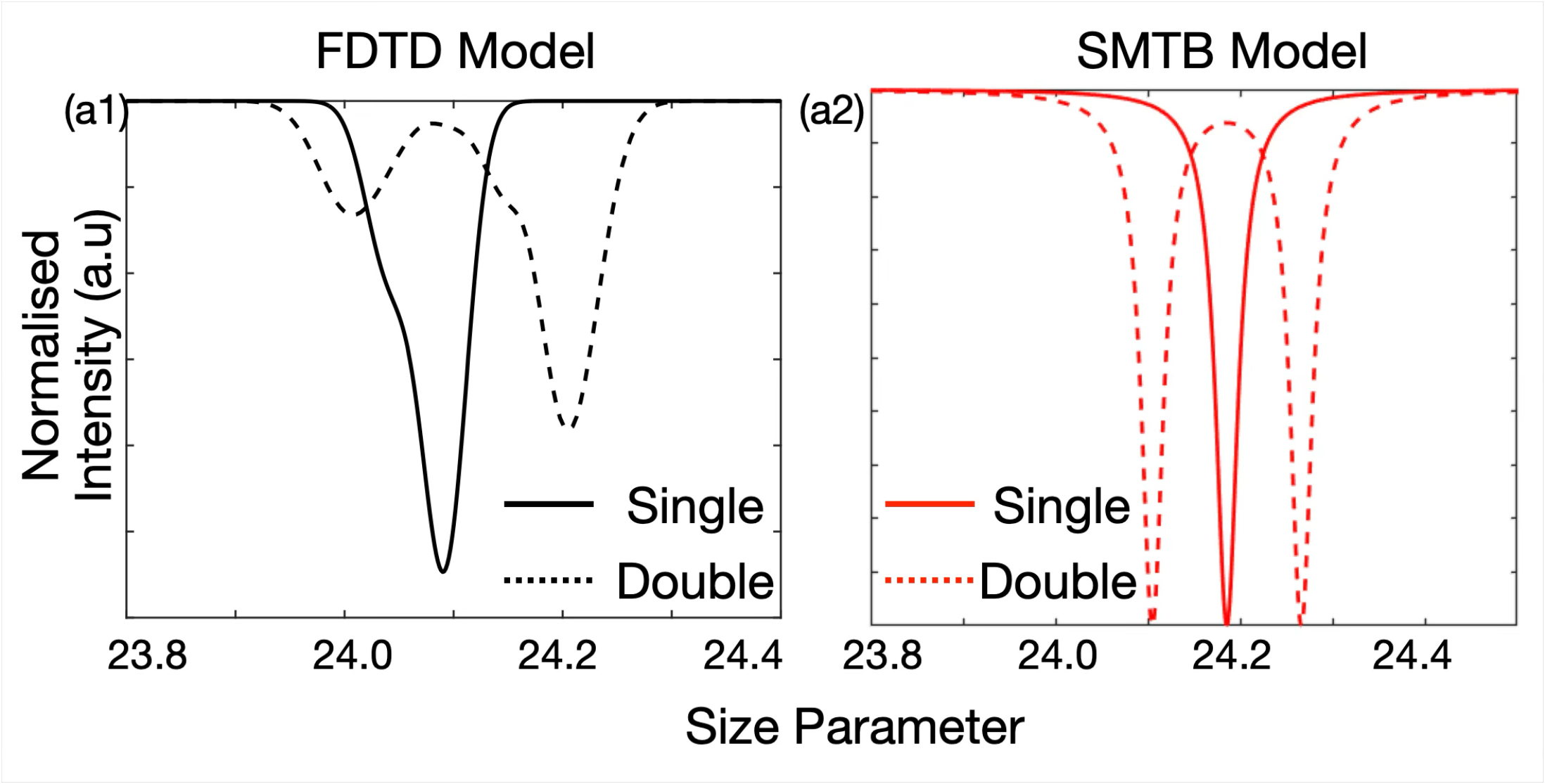}
     \caption{Comparison of normalized intensity vs. the size parameter (= $2\pi a/\lambda$) curves for perovskite coated TiO$_2$ spheres with core diameter 10 $\mu$m and shell thickness 100 nm through (a) FDTD simulations, and (b) SMTB model. The solid curves represent curves for single sphere and the dotted curves state the bi-sphere system.}
     \label{fig:9}
 \end{figure}
 
\begin{figure*}
     \centering
     \includegraphics[scale=0.32]{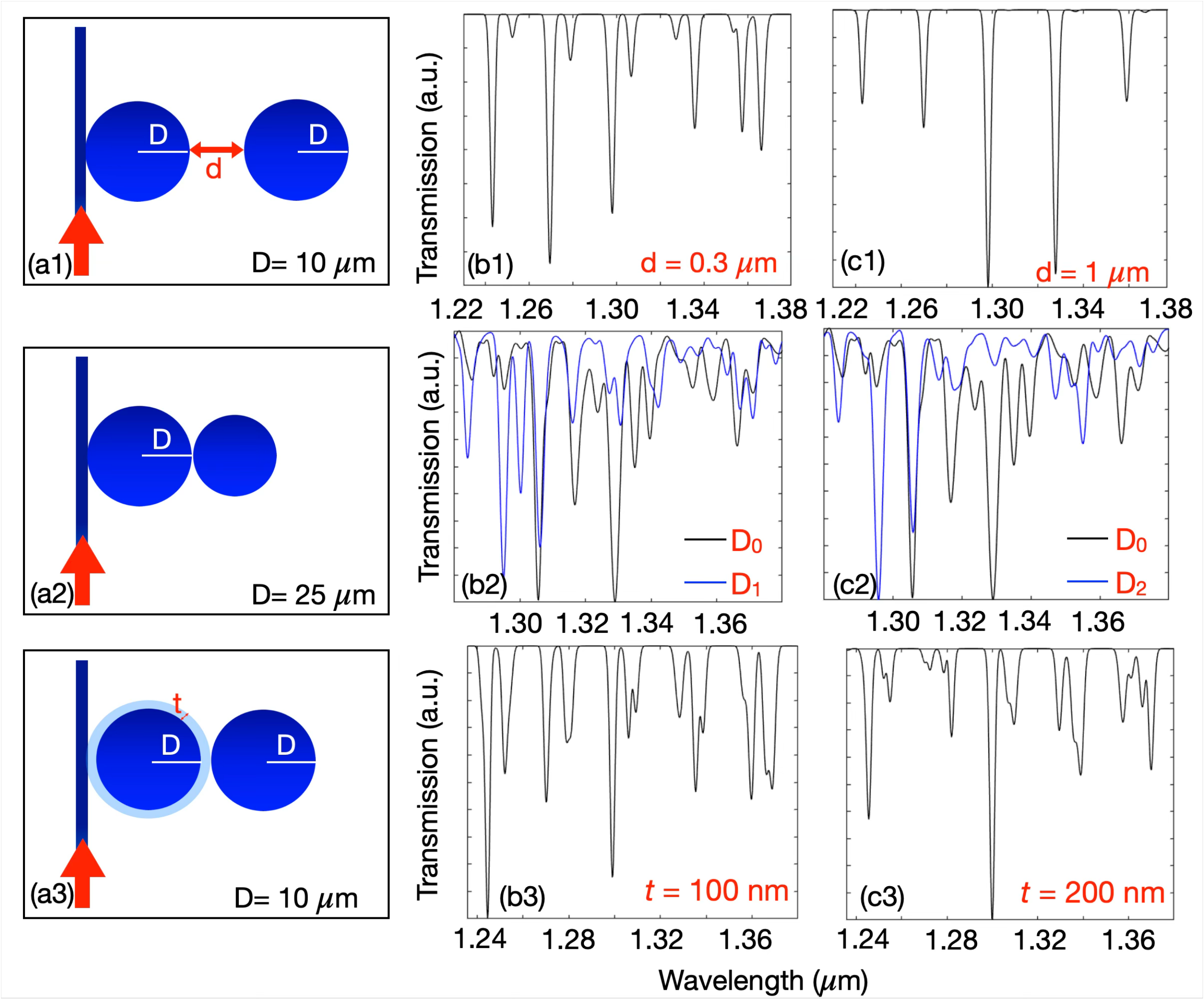}
     \caption{Comparison of sensitivity in a bi-cavity WGM circular resonator. (a1-c1) Separation detuning: increasing the separation `$d$' between the cavities for pristine TiO$_2$ cavities with diameter 10 $\mu m$, (a2-c2) Radius Detuning: coupled TiO$_2$ cavities of different diameter as $D_0$ = 25 $\mu m$ and 25 $\mu m$, $D_1$ = 25 $\mu m$ and 15 $\mu m$, and $D_2$ = 25 $\mu m$ and 7 $\mu m$, and (a3-c3) Refractive index detuning: Coupling between perovskite coated TiO$_2$ cavities and pristine TiO$_2$ cavities on varying the thickness of perovskite shell layer from $t$ = 100 nm to 200 nm, respectively. }
     \label{fig:10}
 \end{figure*}
For core-shell resonators, the normalized spectra obtained from FDTD and tight-binding are shown in Fig.~\ref{fig:9}(a1) and (a2), respectively. The tight-binding model captures the essential splitting behavior, while exhibiting a slight asymmetry relative to the full-wave results. In this case, the operator $S^{(i)}(\omega)$ retains the same diagonal structure in the VSH basis. However, its diagonal elements are given by the layered Mie coefficients derived in Appendix F. No additional coupling channels are introduced, and the Hamiltonian dimension remains unchanged. The presence of the shell modifies only the single cavity scattering response through additional boundary conditions, while the inter-cavity coupling mechanism, governed by the translation operator, remains unchanged.

The asymmetry in the peak positions of the non-degenerate states, as observed in the FDTD spectra (in comparison to the SMTB model) may be attributed to physical effects beyond the minimal tight-binding approximation. In particular, the model neglects radiative self-energy corrections associated with outgoing wave boundary conditions. This leads to small re-normalizations of the complex eigenfrequencies. In addition, the tight-binding formulation assumes a dispersion-less refractive index, whereas the FDTD simulations incorporate the full wavelength-dependent dielectric response of TiO$_2$, including absorption and weak dispersion. Finally, the tight-binding approach retains only the dominant resonant channel for each $(l,m,\beta)$, while the full-wave simulations naturally include contributions from higher-order multipoles and weakly coupled modes. Therefore, additional factors need to be incorporated in the tight-binding model to observe the aforementioned asymmetry. Despite these differences, the relative spectral structure pair formation, ordering, and approximate pair separation of the hybridized modes are reproduced with exceptional consistency, confirming that FDTD and tight-binding both capture the essential coupling in composite dielectric sphere systems.

\section{Detuning parameters for supermode analysis}
In the above the section, with the help of tight-binding formalism, we have described the coupling of resonating cavities. It follows that the hybridization emerges from the interplay of intrinsic scattering response and inter-cavity translation operators. Within this framework, the coupling parameter is derived directly from the underlying WGM resonances and spatial overlap. Therefore, this allows us to quantitatively capture mode splitting, and detuning within a predictive description applicable to both pristine and core–shell systems.

\textbf{Separation detuning:} To further elucidate the role of coupling in dielectric photonic molecules, we have studied the evolution of transmission spectra as a function of inter-cavity separation (d) for pristine ($D$ = 10 $\mu m$) bi-cavity circular resonator molecule (Figure~\ref{fig:10}(a1-c1)). As discussed in the above section, the spectra shows strong mode splitting at zero separation ($d$ = 0), indicating significant hybridization of WGMs due to direct contact coupling. As the separation increases to $d$ = 0.3 $\mu\text{m}$ , the splitting between modes progressively decreases, and the resonances begin to converge toward the uncoupled WGM positions. At larger separations ($d$ = 1 $\mu\text{m}$), the modes become nearly indistinguishable from those of isolated resonators, confirming that the coupling is primarily mediated by evanescent overlap, which decays rapidly with distance.


\textbf{Diameter detuning:} To examine the dependence of coupling on cavity size, we consider three bi-cavity systems with increasing diameter mismatch: $D_0$ (25 - 25 $\mu m$), $D_1$ (25 - 15 $\mu m$), and $D_2$ (25 - 7 $\mu m$). As shown in Figure~\ref{fig:10}(a2-c2), the set of hybridized modes remains spectrally identifiable across all configurations, while their relative excitation strengths are strongly modified. This reflects a redistribution of the electromagnetic field between the cavities, which alters the overlap with the excitation channel.

In the symmetric case for $D_0$ (shown as black lines in Figure~\ref{fig:10}(b2-c2)), modes are uniformly excited, producing well-defined resonance dips. With moderate mismatch ($D_1$) in Figure~\ref{fig:10}(b2), detuning becomes comparable to the coupling strength, leading to partially hybridized modes with asymmetric field distribution and reduced excitation efficiency. For larger mismatch ($D_2$) in Figure~\ref{fig:10}(c2), detuning dominates over coupling, suppressing hybridization and localizing modes within individual cavities. Modes confined to the larger cavity, which couples more efficiently to the input field, exhibit enhanced excitation and deeper transmission dips. Thus, while the modal structure is preserved, diameter mismatch provides a direct route to selectively control mode visibility and excitation strength.

\textbf{Refractive index detuning:} To isolate the role of material detuning, we compare the transmission spectra of pristine TiO$_2$ cavities with their perovskite-coated counterparts (Figure~\ref{fig:10}(a3)). Increasing the coating thickness from $t$ = 100 nm to 200 nm leads to a systematic reduction in mode splitting and a redistribution of resonance intensities Figure~\ref{fig:10}(b3 and c3). This reflects a transition from a coupling-dominated regime in pristine structures to a detuning-dominated regime in coated systems, where material-induced modifications to the dielectric environment suppress inter-cavity interaction.

Despite this transition, the underlying supermode structure is preserved, indicating that hybridization remains robust while its strength is tunable. The coating thus enables control over spectral spacing and mode density without altering geometry. Furthermore, the resulting non-uniform mode excitation provides an additional degree of freedom for sensing, where both resonance shifts and intensity variations contribute to enhanced spectral selectivity and information content.

\section{Conclusion}
We have investigated WGM hybridization in photonic molecules composed of pristine and perovskite-coated TiO$_2$ microresonators, combining full-wave FDTD simulations with a single-mode tight-binding model. The central outcome of this work is that the spectral organization of high-index photonic molecules can be controlled through two complementary and independently tunable parameters: the connectivity of the resonator network and the dielectric environment surrounding each resonator. Topology determines the symmetry, multiplicity, and relative organization of the collective supermodes, whereas dielectric-shell engineering modifies the strength and distribution of their spectral splitting without requiring a change in molecular connectivity. In this sense, the perovskite shell provides a material degree of freedom that complements the geometry-defined coupling pathways.

Our analysis of linear chains and compact clusters have demonstrated that each configuration exhibits distinct and reproducible spectral fingerprints that persist across variations in cavity size and material composition. Representative hybridized resonances exhibit $Q$-factors of order \(10^3\) and remain spectrally resolved, allowing the coupling-induced spectral structure to be clearly identified. The introduction of perovskite coatings provides diameter- and thickness-dependent control of the coupling strength and spectral splitting, with coupling enhancement in the smaller-diameter regime and suppression at larger diameters. Despite these changes, the topology-dependent spectral structure remains intact, indicating that hybridization can be tuned independently of geometry without significant degradation of resonant quality. Further, the single-mode tight-binding formulation provides a consistent description of the observed behavior, while reproducing the splitting of modes, and establishing a direct correspondence between physical system parameters and the effective coupling parameters.

Overall, the results presented in this work establish photonic molecules as tunable platforms in which topology and material composition act as control parameters for engineering hybridized optical states. This provides a general framework for designing coupled resonator systems with controllable spectral response, with potential applications in sensing, spectral filtering, and hybrid photonic-optoelectronic devices.

\section*{Computational details}
All electromagnetic simulations were performed using a FDTD method implemented in Lumerical Solutions \cite{anys}. The simulations employ a 2-D cross-sectional representation of the microspheres as circular resonators and rectangular waveguide representation of the tapered-fiber excitation. The dielectric pristine TiO$_2$ as well as CsPbBr$_3$ coated TiO$_2$ microcavities placed in close contact with the waveguide (typically 100 nm for simulation), as shown Figure~\ref{fig:2}. The material dispersion of TiO$_2$ and CsPbBr$_3$ has been included from the reference \cite{Cui_2016} and \cite{Brennan2024}, respectively.

A fundamental TM mode source has been launched within the 1.5 $\mu$m waveguide ($n$=1.45) from bottom upwards to study the evanescent coupling into the WGMs of the cavities. The guided mode has then been computed using the built-in eigenmode solver. This configuration provided efficient phase matched coupling into the circulating modes of the microsphere cavities. Transmission spectra were collected using a power monitor placed at the waveguide output and normalized using a reference (no-sphere) simulation. Perfectly matched layer (PML) absorbing boundary conditions were applied along the $x$ and $y$ directions of the computational domain. A nonuniform meshing strategy with a sub-nanometer mesh override region surrounding the sphere waveguide interface has been used to resolve the highly confined WGM fields. For each pristine and core–shell resonator configuration, the corresponding isolated-cavity resonance was first determined and used as the reference wavelength `$\lambda_o$' for the subsequent normalized-detuning analysis of the coupled photonic molecules.
        
\section*{Acknowledgments}
The authors are thankful for the mathematical insights from Dr. Hareram Swain and discussions with Prof. Sudakar Chandran. Also, this work was funded by Department of Science and Technology (DST)-India, for funding through grant number DST/TMD/IC-MAP/2K20/03(C).

\section*{Conflict of interest}
The authors declare no conflict of interests.  

\newpage 

\appendix

\section{Transmission spectra for longer photonic molecules chain}
Figure~\ref{fig:A1} shows the orientation of photonic molecules and the splitting of modes in the transmission spectra for linear chains of $N$=5 (Figure~\ref{fig:A1}(a)) and $N$=6 (Figure~\ref{fig:A1}(b)). 

\section{$Q$-factor analysis}
The quality factors of the hybridized resonances were extracted from the simulated transmission spectra using Lorentzian fitting, as shown in Figure~\ref{fig:A8}. For an individual transmission dip \cite{Peng2014}, the spectral response was described by:
\begin{equation}
    T(\lambda) = C - \frac{A_1}{1+(\frac{\lambda-\lambda_\mathrm{res_1}}{w_1})^2} + \frac{A_2}{1+(\frac{\lambda-\lambda_\mathrm{res_2}}{w_2})^2}
\end{equation}
where $C$ is the baseline, $A_1$, $A_2$ are the resonance amplitudes and $\lambda_\mathrm{res_1}$, $\lambda_\mathrm{res_2}$ are the fitted resonance wavelength with $w$ as the half-width at half-maximum of hybridized supermodes. The resonance linewidth was obtained as $\Delta$$\lambda$=2$w$, and the spectrally extracted quality factor was calculated as: 
\begin{equation}
    Q = \frac{\lambda_\mathrm{res}}{\Delta \lambda_\mathrm{FWHM}}
\end{equation}
The values reported in \ref{tab:I} and \ref{tab:II} are spectral $Q$-factors extracted from the linewidths of the simulated transmission resonances. These values represent the effective spectral $Q$-factors of the hybridized resonances and are not separately decomposed into intrinsic, radiative, absorptive, or external-coupling contributions.

\section{Coupling and detuning in perovskite coated photonic molecules}
Figure~\ref{fig:A2} and Figure~\ref{fig:A3} show the normalized detuning of hybridized supermodes for coated photonic molecules with shell thicknesses of 100 nm and 200 nm, respectively. In both cases, topology-dependent patterns are preserved, with linear chains showing symmetric spacing and compact clusters exhibiting asymmetric distributions. Increasing the coating thickness leads to compression of the detuning range, indicating reduced inter-cavity coupling.

Figure~\ref{fig:A4} presents the diameter dependence of coupling strength for pristine and coated systems. The coated structures show a non-monotonic variation in coupling relative to pristine cavities, reflecting the interplay between evanescent field overlap at small diameters and increased confinement and dielectric separation at larger sizes. Together, these results confirm that coating primarily modifies the strength of hybridization while preserving the underlying topology-dependent spectral structure. 

\begin{figure} [htbp]
    \centering
    \includegraphics[scale=0.27]{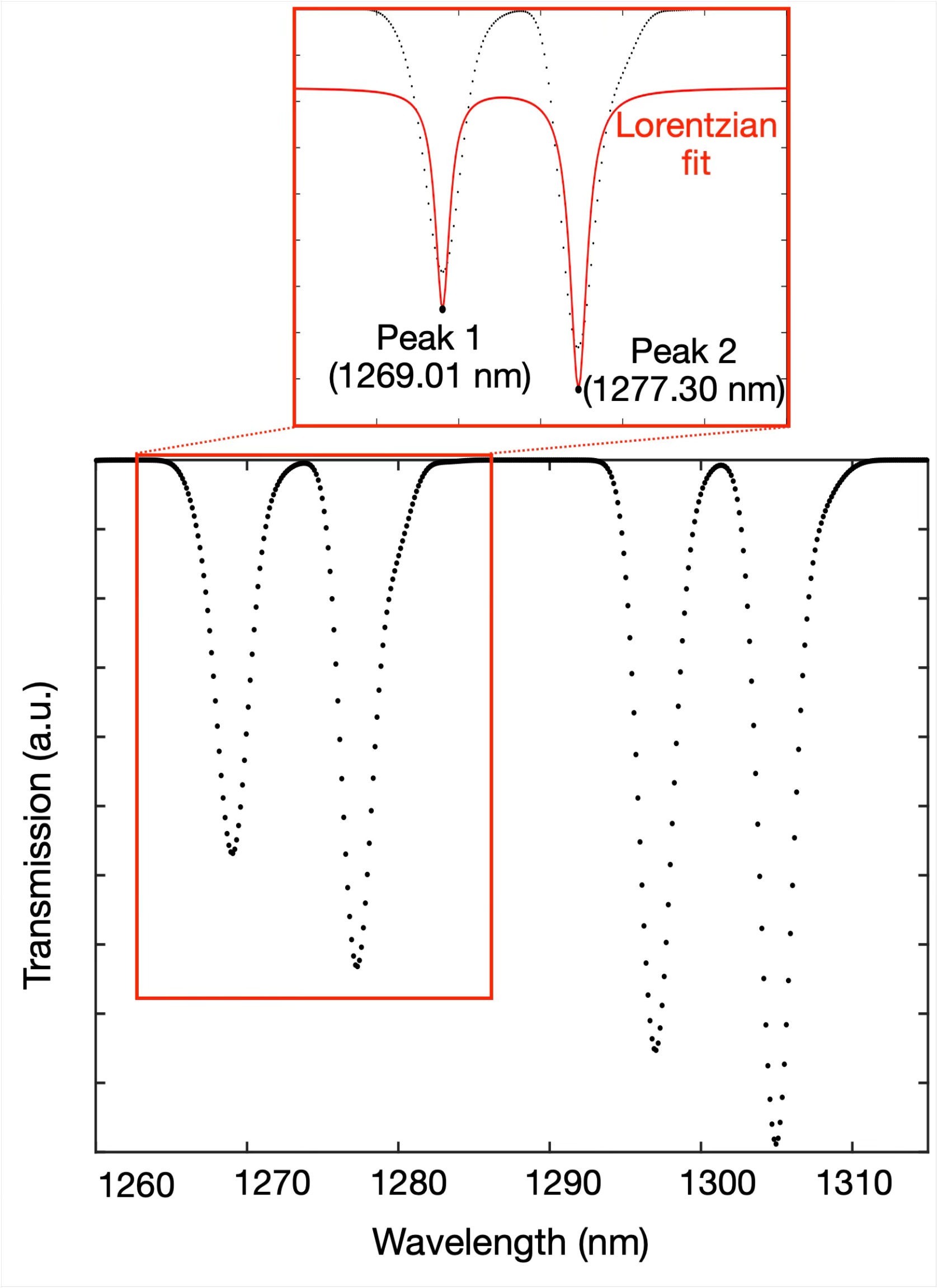}
    \caption{Transmission spectra of bi-sphere pristine system, with the inset showing the normalized transmission spectrum (dotted black line) and corresponding two-peak Lorentzian fit (solid red line) used to extract the resonance wavelengths and full widths at half maximum of the hybridized modes. The selected resonances were fitted using a multi-Lorentzian function with an adjustable baseline, and the resonance wavelengths and FWHMs obtained from the fits were used to calculate the corresponding $Q$-factor (\(Q=\lambda_0/\Delta\lambda_{\mathrm{FWHM}}\)).}
    \label{fig:A8}
\end{figure}

\begin{figure*} [htbp]
    \centering
    \includegraphics[scale=0.3]{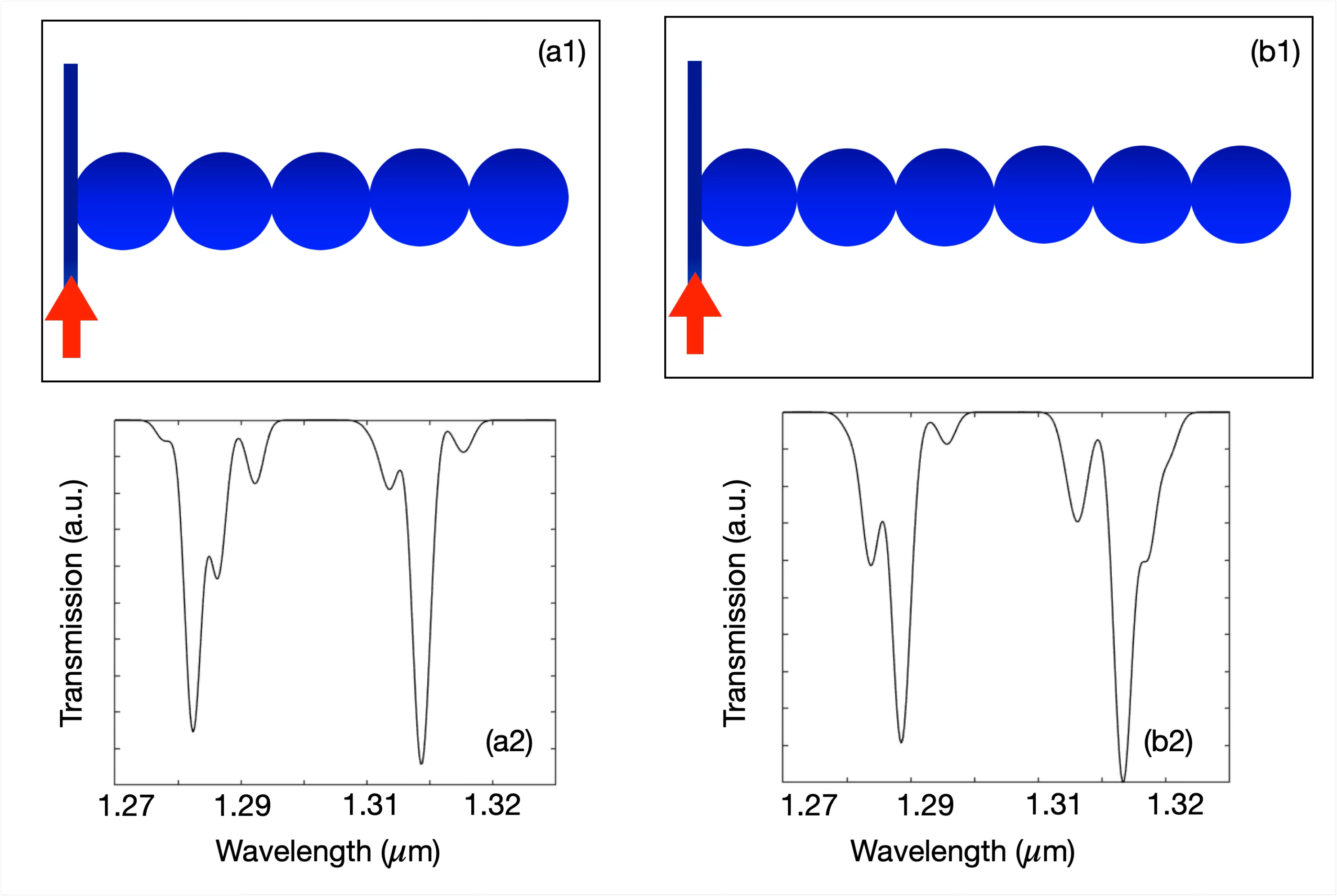}
    \caption{Simulated transmission spectra of hybridization of WGM in 10 $\mu$m TiO$_2$ photonic molecules. (a1 and a2) Linear photonic molecular configuration for respective $N$ = 5 and 6, and the corresponding transmission spectra (a2 and b2).}
    \label{fig:A1}
\end{figure*}

\begin{figure} [htbp]
    \centering
    \includegraphics[scale=0.25]{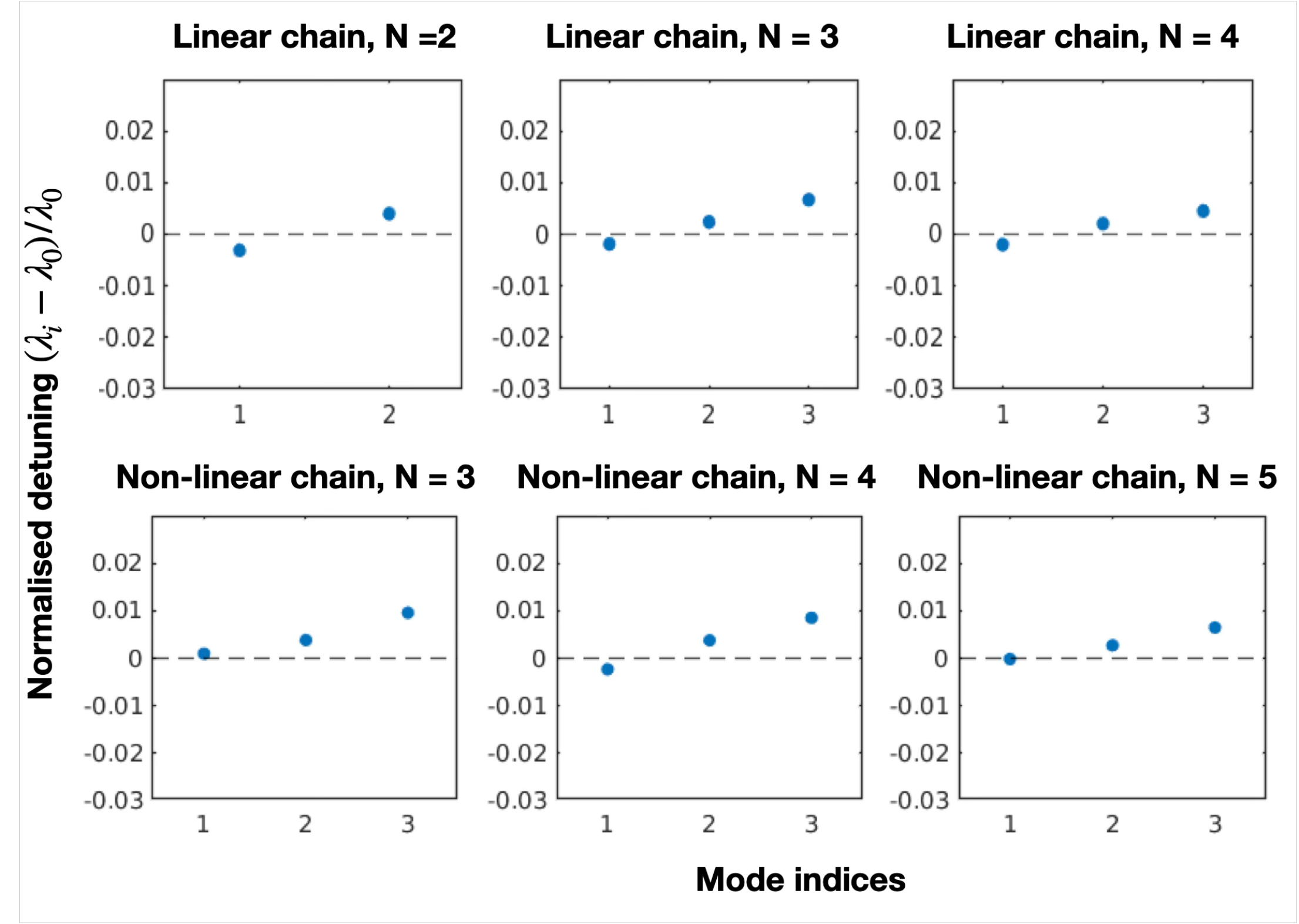}
    \caption{Normalized detuning in the hybridized supermodes relative to the corresponding isolated-cavity parent resonance, for linear and compact TiO$_2$ photonic molecular systems with core diameter at D=10 $\mu$m and perovskite shell thickness at 100 nm.}
    \label{fig:A2}
\end{figure}
\begin{figure} [htbp]
    \centering
    \includegraphics[scale=0.25]{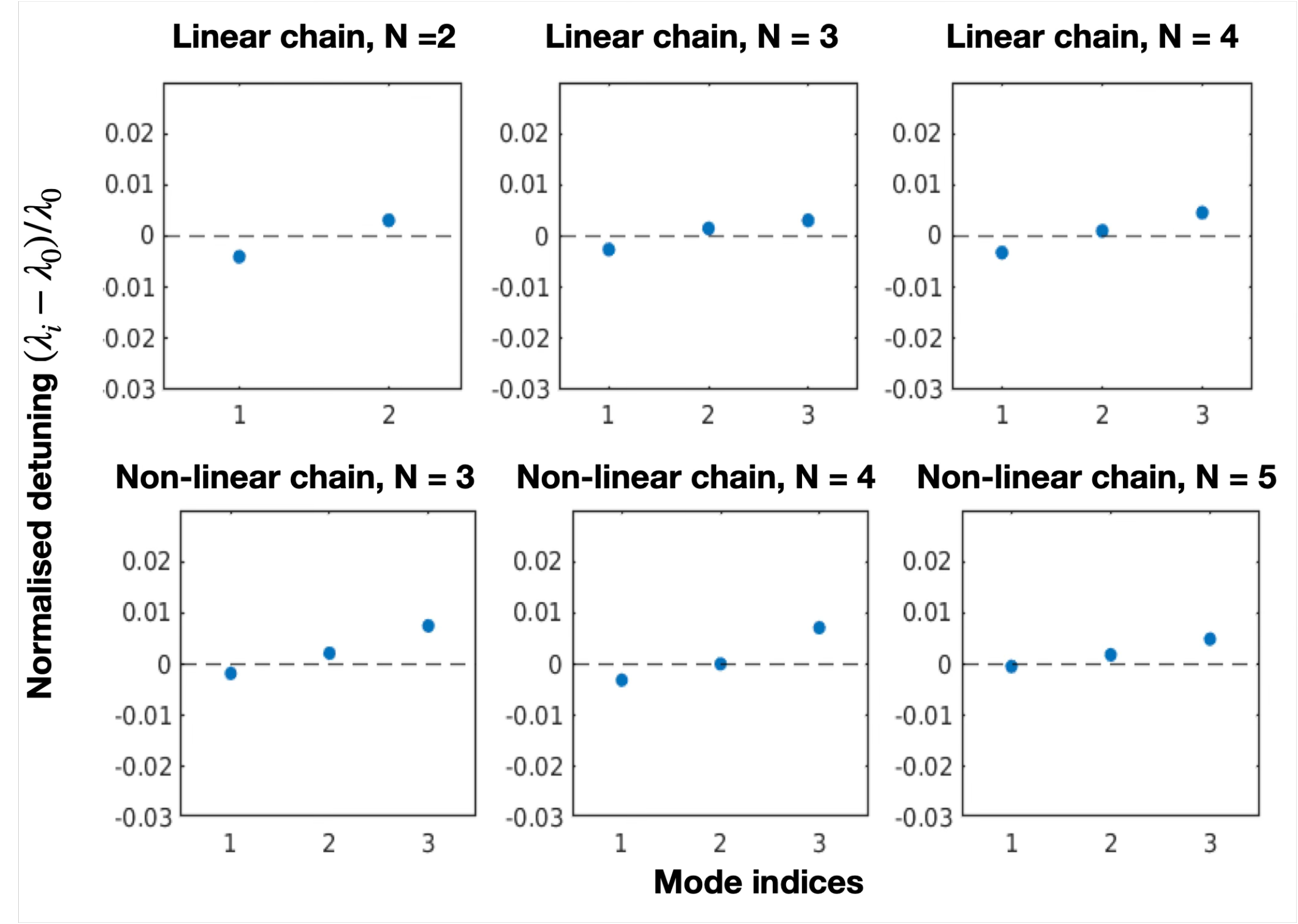}
    \caption{Normalized detuning in the hybridized supermodes relative to the corresponding isolated-cavity parent resonance, for linear and compact TiO$_2$ photonic molecular systems with core diameter at D=10 $\mu$m and perovskite shell thickness at 200 nm.}
    \label{fig:A3}
\end{figure}
\begin{figure*} [htbp]
    \centering
    \includegraphics[scale=0.32]{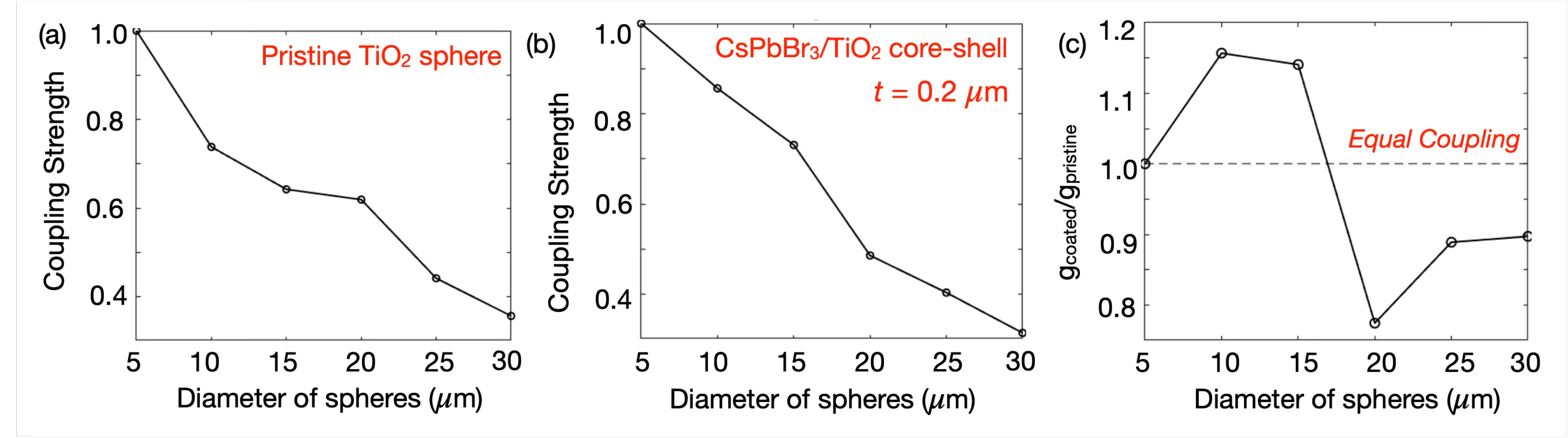}
    \caption{Diameter dependent coupling strength for (a) pristine and (b) perovskite-coated cavities with perovskite thickness $t$ = 200 $nm$, and (c) Effective coupling of both the configurations in bi-cavity circular resonator system.}
    \label{fig:A4}
\end{figure*}
\begin{figure*} [htbp]
    \centering
    \includegraphics[scale=0.32]{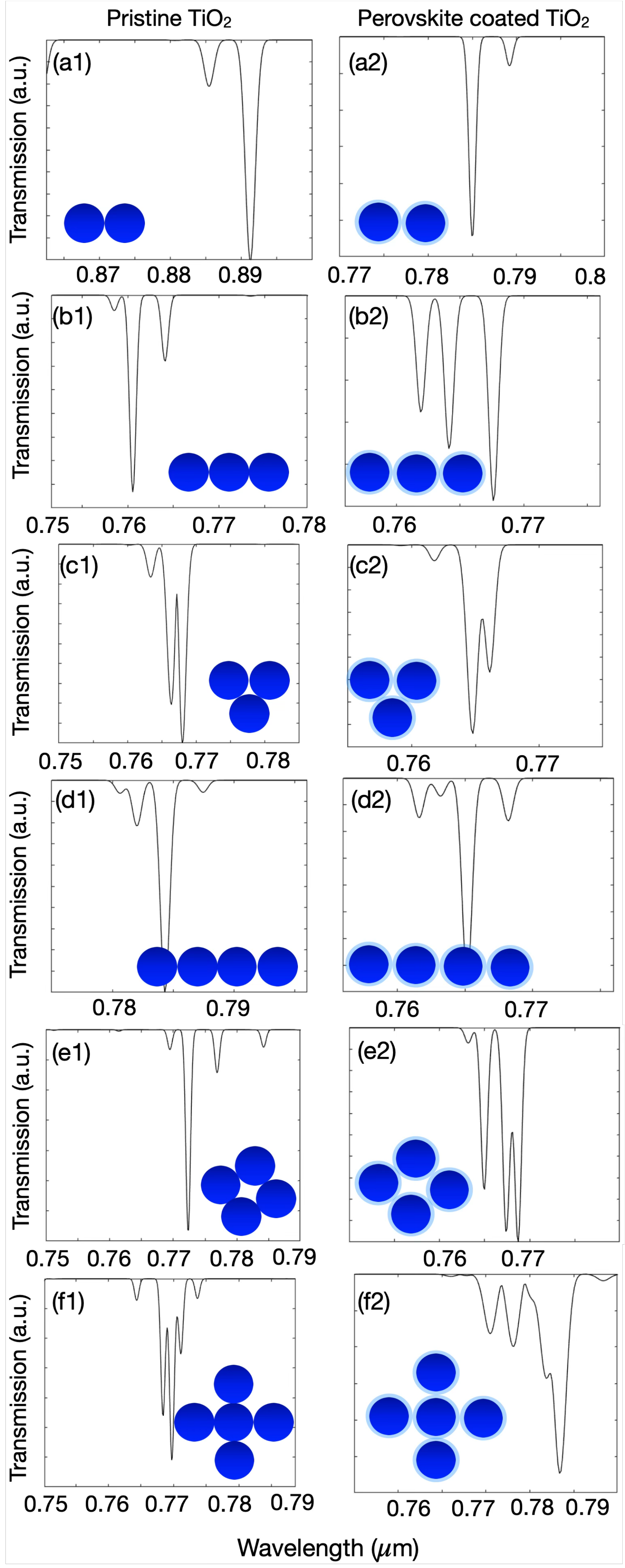}
    \caption{Transmission spectra of pristine and coated photonic molecules for reduced cavity diameter of (a1-f1) pristine TiO$_2$ spheres with diameter 4 $\mu$m, and (a2-b2) perovskite coated TiO$_2$ spheres, with shell thickness of 100 nm. The inset shows the photonic molecular arrangement considered for calculating the respective transmission spectra. Stronger evanescent overlap at smaller sizes leads to clearer resolution of hybridized supermodes, providing enhanced visualization of mode splitting and spectral signatures across configurations.}
    \label{fig:A5}
\end{figure*}
\begin{figure*} [htbp]
    \centering
    \includegraphics[scale=0.28]{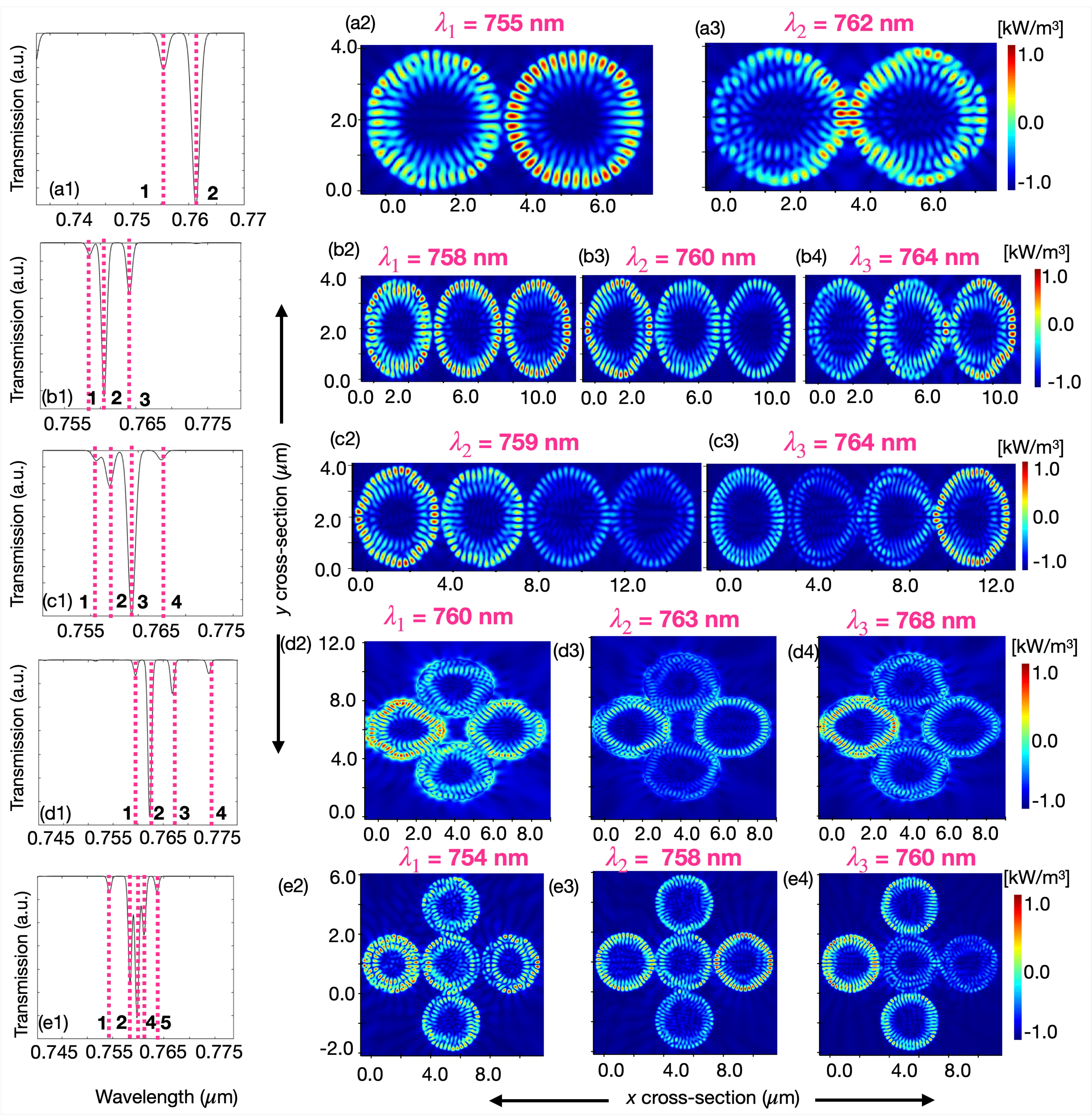}
    \caption{The electric-field distribution profiles of the photonic molecular systems calculated at the wavelengths corresponding to different supermodes for pristine TiO$2$ spheres with diameter 4 $\mu$m. The following molecules are represented: linear chain formed by (a1) two circular, (b1) three circular and (c1) four circular resonators; (d1) quadrumer resonator, and (e1) five-cross circular resonator. Simulated EM ﬁeld maps at different supermode eigen wavelengths are presented for each molecule in (a2-a3), (b2-b4), (c2-c3), (d2-d4), and (e2-e4) respectively.}
    \label{fig:A6}
\end{figure*}
\begin{figure*} [htbp]
    \centering
    \includegraphics[scale=0.28]{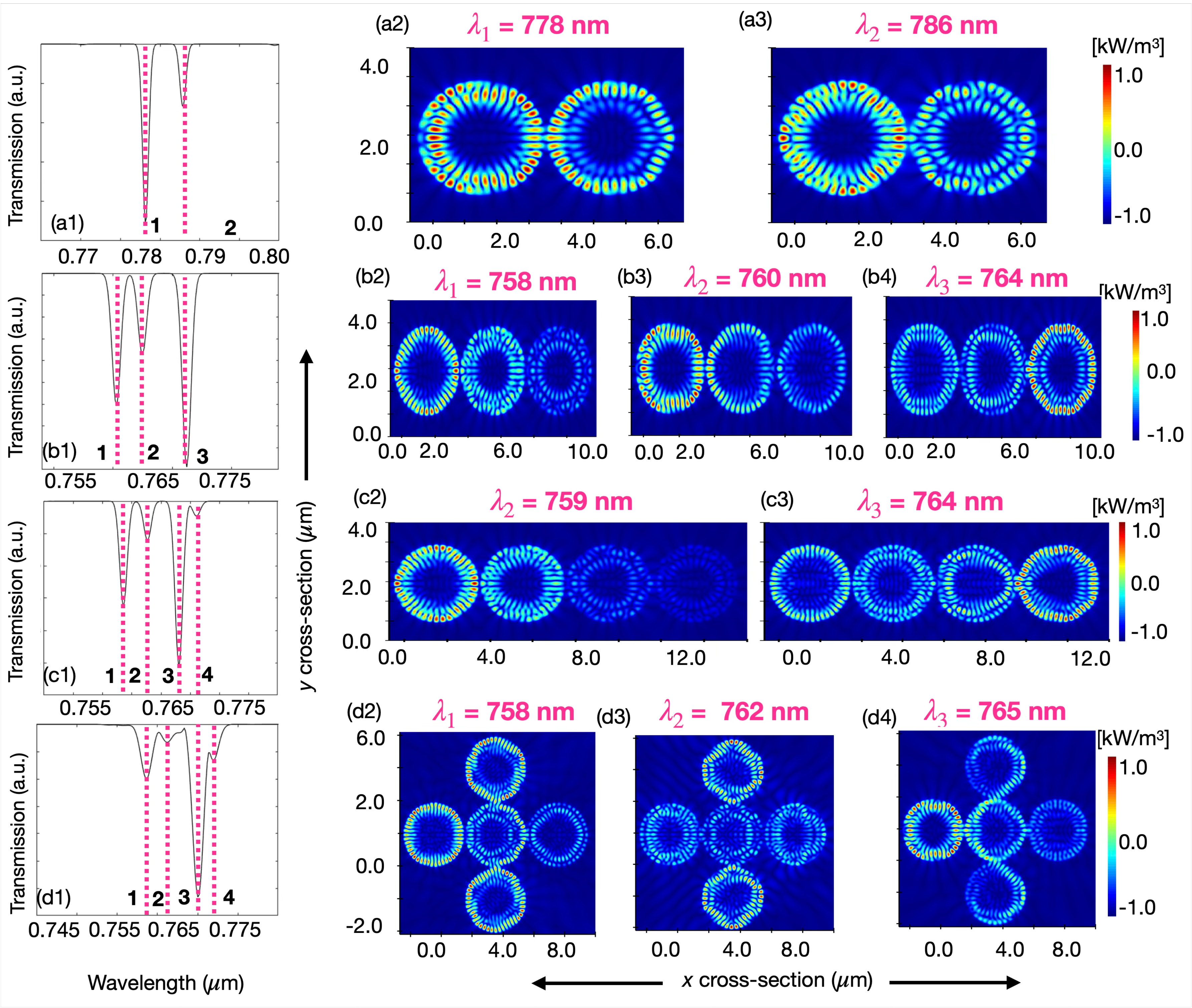}
    \caption{The electric-field distribution profiles of the photonic molecular systems calculated at the wavelengths corresponding to different supermodes for core-shell perovskite coated TiO$_2$ spheres with core diameter 4 $\mu$m and perovskite thickness 100 nm. The following molecules are represented: linear chain formed by (a1) two circular, (b1) three circular and (c1) four circular resonators; and (d1) five-cross circular resonator. Simulated EM ﬁeld maps at different supermode eigen wavelengths are presented for each molecule in (a2-a3), (b2-b4), (c2-c3), and (d2-d4) respectively.}
    \label{fig:A7}
\end{figure*}

\section{Hybridization of supermodes in smaller diameter cavities}
Figure~\ref{fig:A5}-~\ref{fig:A7} present the transmission spectra and corresponding field distributions of pristine and perovskite-coated photonic molecules for reduced cavity diameter (D = 4 $\mu$m), chosen to enable clearer visualization of hybridized modes. At larger diameters (10–25 $\mu$m), WGMs are strongly confined near the resonator surface, resulting in closely spaced spectral features and less distinguishable field distributions across supermodes. In contrast, smaller cavities exhibit increased evanescent field extension, leading to stronger effective inter-cavity interaction and improved resolution of hybridized states. As shown in Figure~\ref{fig:A5}, the transmission spectra display well-resolved mode splitting that preserves the characteristic topology-dependent patterns observed in the main text, including symmetric distributions in linear chains and broadened, asymmetric features in compact clusters. The corresponding field distributions in  Figure~\ref{fig:A6} (pristine) and  Figure~\ref{fig:A7} (coated) further illustrate the evolution of hybridized modes, ranging from delocalized to progressively localized states. In coated systems, reduced field overlap at the inter-cavity regions and enhanced localization are observed, consistent with the combined effect of reduced coupling strength and material-induced detuning. Despite the change in cavity size, the qualitative features of hybridization, including mode symmetry, spatial redistribution, and topology dependence, remain consistent, confirming that the reduced-diameter systems provide a representative and physically meaningful visualization of the underlying coupling mechanisms. \\

\section{Vector Spherical Harmonics}
Here we construct the internal and scattered field bases $E_{lm}^{\beta}$ from first principles, starting with scalar spherical harmonics and proceeding through Maxwell’s eigen functions in spherical geometry. The scalar spherical harmonics, which form the complete orthogonal basis for all angular functions on the sphere are given as:
\begin{align}
Y_{lm}(\theta,\phi)=P_l^m(\cos\theta)e^{im\phi},
\end{align}
where $P_l^m$ are the associated Legendre polynomials.
These functions satisfy the well known orthogonality relation given as:
\begin{align}
\int_{0}^{2\pi}\int_{0}^{\pi}Y_{lm}(\theta,\phi)Y_{pq}^{*}(\theta,\phi)\sin\theta d\theta d\phi = \delta_{lp}\delta_{mq},
\end{align}
This condition is necessary such that the multipole channels $(l,m)$ remain uncoupled in the single sphere response.
The vector solutions of Maxwell’s equations is constructed from the angular momentum operator, given by $L=-ir\times\nabla$, to the scalar harmonics.
This defines the basic angular vector spherical harmonics as:
\begin{align}
 X_{lm}(\theta,\phi)=\frac{1}{\sqrt{l(l+1)}}LY_{lm}  \\ \nonumber
= \frac{1}{\sqrt{l(l+1)}}(i\pi_l^m(\theta)e^{im\phi}\hat{\theta}-\tau_l^m(\theta)e^{im\phi}\hat{\phi})
\end{align}
where the angular functions are: $\pi_l^m(\theta)=\frac{m}{\sin\theta}P_l^m(\cos\theta)$ and $\tau_l^m(\theta)=\frac{dP_l^m(\cos\theta)}{d\theta}$, contain all dependence on the associated Legendre polynomials. This derivation confirms the explicit dependence of $X_{lm}$ on purely tangential components, rather the radial components.

Further, the solutions to the vector Helmholtz equation ($\nabla^2\mathbb{E}$+$k^2\mathbb{E}=0$), can be constructed using the angular harmonic together with a radial function ($f_l(kr)$). The solutions for TE and TM modes are given as:
\begin{subequations}
    \begin{widetext}
    \begin{align}
      E_{lm}^{M}(f_l,kr,\theta,\phi) = f_l(kr)X_{lm}(\theta,\phi)= f_l(kr)\frac{1}{\sqrt{l(l+1)}}\left[i\pi_l^m(\theta)e^{im\phi}\hat{\theta}-\tau_l^m(\theta)e^{im\phi}\hat{\phi}\right] \\
      E_{lm}^{N}(f_l,kr,\theta,\phi) = \frac{1}{k}\nabla\times\left[f_l(kr)X_{lm}(\theta,\phi)\right] = \frac{l(l+1)}{k r}f_l(kr)P_l^m(\cos\theta)e^{im\phi}+ \\ \nonumber
      \frac{1}{k r}\frac{d}{dr}\left(rf_l(kr)\right)\tau_l^m(\theta)e^{im\phi}+ 
       \frac{1}{k r}\frac{d}{dr}\left(rf_l(kr)\right)\pi_l^m(\theta)e^{im\phi}
    \end{align}
\end{widetext}
\end{subequations}

The equations state that the TE mode contain no radial component, while the TM modes contain all of the three components (radial, polar as well as azimuthal). Since, the angular part of each mode is ultimately constructed from $Y_{lm}$, the Maxwell eigenfunctions inherit the orthogonality, given as:
\begin{align}
    \int E_{lm}^{\beta}(f_l)\cdot E_{pq}^{\beta*}(f_p)\sin\theta d\theta d\phi=
\delta_{lp}\delta_{mq} \mathcal{N}_l^{\beta}(r), \\ \nonumber
\qquad \beta\in{M,N}
\end{align}
The normalization factor ($\mathcal{N}_{l}^{\beta}(r)$) depends only on the radial functions and plays no role in the coupling formulation.

\section{Multipole Coefficients for a Single Pristine Sphere}
The fields associated with each sphere are expanded in the VSH basis with $\nu \equiv (l,m,\beta)$, $\beta\in{M,N}$. For a sphere centered at ($R_i$), the local coordinates are defined as $r_i = r - R_i$. The internal field is expressed as:
\begin{align}
  \mathbf{E}^{(i)}_{\mathrm{int}}(r_i)= \sum_{\nu}a^{(i)}_{\nu}E^{(i)}_{\nu}(j_l, k_i r_i), 
\end{align}
where $a^{(i)}_{\nu}$ are the internal multipole coefficients and $E_{\nu}$ are regular VSHs constructed from the spherical Bessel functions ($j_l$). The scattered field is expressed as:
\begin{align}
\mathbf{E}^{(i)}_{\mathrm{sca}}(r_i)= \sum_{\nu}b^{(i)}_{\nu}E^{(i)}_{\nu}(h^{(1)}_l, k_or_i), 
\end{align}
where $b^{(i)}_{\nu}$ are the outgoing multipole coefficients and $E_{\nu}$ are radiating VSHs based on spherical Hankel functions $h^{(1)}_l$.
The incident field (external illumination + scattered field from the partner sphere) is expanded as:
\begin{align}
\mathbf{E}^{(i)}_{\mathrm{inc}}(r_i)=\sum_{\nu}c^{(i)}_{\nu}E^{(i)}_{\nu}(j_l, k_o r_i),
\end{align}
where $c^{(i)}_{\nu}$ are the incident-field multipole coefficients.

The boundary conditions on the surface of sphere $i$ relate these coefficients through two diagonal operators. The coefficients $a^{(i)}_{\nu}$ describe the amplitudes of the regular (Bessel-type) VSH components inside sphere $i$. These amplitudes are determined by the total incident field at the sphere surface, whose VSH expansion coefficients are denoted $c^{(i)}_{\nu}$.
Enforcing continuity of the tangential electric field at ($r_i = a_i$) yields a diagonal linear relation between $a^{(i)}_{\nu}$ and $c^{(i)}_{\nu}$ as : 
\begin{align}
  a^{(i)}_{\nu}=\sum_{\nu'} R^{(i)}_{\nu\nu'}(\omega)c^{(i)}_{\nu'},
\end{align}
where $R^{(i)}_{\nu\nu'}$ is the single sphere internal response operator. 
\begin{subequations} \label{eq:21}
\begin{align}
R^{(i)}_{\nu\nu'}(\omega)=\delta_{\nu\nu'}R_l^{(i)}(\omega),
\qquad
\label{eq:21a} \\[6pt]
R_l^{(i)}(\omega) \approx \frac{j_l(k a_i)}{j_l(n_i k a_i)}
\label{eq:21b}
\end{align}
\end{subequations}
The quantity $R_l^{(i)}(\omega)$ therefore represents the ratio of the internal to external regular spherical Bessel components for a given multipole order ($l$).

The scattered field coefficients $b^{(i)}_{\nu}$ represent the amplitudes of the outgoing spherical Hankel components radiated by sphere $i$. Imposing the continuity of both tangential electric and magnetic fields at the sphere boundary yields a linear mapping from the internal coefficients $a^{(i)}_{\nu}$ to the scattered coefficients $b^{(i)}_{\nu}$,

\begin{subequations} \label{eq:21}
\begin{align}
b^{(i)}_{\nu}=\sum_{\nu'} S^{(i)}_{\nu\nu'}(\omega)a^{(i)}_{\nu'},
\label{eq:21a} \\[6pt]
S^{(i)}_{\nu\nu'}(\omega)=\delta_{\nu\nu'}S_l^{(i)}(\omega)
\label{eq:21b}
\end{align}
\end{subequations}

where $S^{(i)}(\omega)$ is the Mie scattering operator of the isolated sphere. This operator is diagonal in the VSH basis, and its diagonal entries $S_l^{(i)}(\omega)$ correspond exactly to the standard Mie scattering coefficients for TE ($\beta = M$) and TM ($\beta = N$) modes, given as:
\begin{subequations} \label{eq:Mie scattering in prsitine spheres}
\begin{align}
S_{l}^{(i),M}(\omega)=  \frac{m\psi_l(mx)\psi_l'(x)-\psi_l(x)\psi_l'(mx)}{m\psi_l(mx)\xi_l'(x)-\xi_l(x)\psi_l'(mx)}
\label{eq:21a} \\[6pt]
S_{l}^{(i),N}(\omega)=  \frac{\psi_l(mx)\psi_l'(x)-m\psi_l(x)\psi_l'(mx)}{\psi_l(mx)\xi_l'(x)-m\xi_l(x)\psi_l'(mx)}
\label{eq:21b}
\end{align}
\end{subequations}
where, $mx = n_ika_i$, with $a_i$ as the radius of sphere, $k$is the exterior wavenumber, and $m$ is the refractive index of the sphere relative to the background. The functions $\psi_l(z) = z j_l(z)$ and $\xi_l(z) = z h_l^{(1)}(z)$ are Riccati-bessel functions.

\section{Multipole Coefficients for a Core-Shell Sphere}
For a core-shell spherical resonator, the presence of two interfaces requires the field expansion to be defined separately in the core, shell, and exterior regions. The fields are expanded in the VSH basis in core ($r_i < a_1$), shell ($a_1 < r_i < a_2$) and external region ($r_i > a_2)$ is given respectively, as:
\begin{subequations}\label{Eq:Boundary conditions_core-shell}
\begin{align}
& \mathbf{E}^{(i)}_{core}(r_i)= \sum_{\nu} a^{(i)}_{\nu}E^{(i)}_{\nu}(j_l, k_1 r_i), \\
& \mathbf{E}^{(i)}_{\mathrm{shell}}(r_i)= \sum_{\nu}\left[\alpha^{(i)}_{\nu}E^{(i)}_{\nu}(j_l, k_2 r_i) +\beta^{(i)}_{\nu} E^{(i)}_{\nu}(y_l, k_2 r_i)\right], \\
& \mathbf{E}^{(i)}_{\mathrm{sca}}(r_i)= \sum_{\nu} b^{(i)}_{\nu} E^{(i)}_{\nu}(h_l^{(1)}, k_o r_i).
\end{align}
\end{subequations}
where $a^{(i)}_{\nu}$ and $b^{(i)}_{\nu}$ are the multipole coefficients VSHs constructed from spherical Bessel functions and Hankel functions, and $\alpha^{(i)}_{\nu}$ and $\beta^{(i)}_{\nu}$ represent the coefficients of the regular solutions in the shell region. Further, as discussed in the Equations \ref{eq:Mie scattering in prsitine spheres}, the diagonal entries $S_l^{(i)}(\omega)$ for TE ($\beta = M$) and TM ($\beta = N$) modes in isolated core-shell spheres takes the form:
\begin{widetext}
\begin{subequations} \label{eq:Mie scattering in core-shell spheres}
\begin{align} 
&S_l^{(i),M}(\omega)=\frac{m_2 \psi_l'(x)\left[\psi_l(m_2 x)-A_l \chi_l(m_2 x)\right]-\psi_l(x)\left[\psi_l'(m_2 x)-A_l \chi_l'(m_2 x)\right]}{m_2 \xi_l'(x)\left[\psi_l(m_2 x)-A_l \chi_l(m_2 x)\right]-\xi_l(x)\left[\psi_l'(m_2 x)-A_l \chi_l'(m_2 x)\right]} \\
& S_l^{(i),N}(\omega)=\frac{\psi_l'(x)\left[\psi_l(m_2 x)-A_l \chi_l(m_2 x)\right]-m_2\psi_l(x)\left[\psi_l'(m_2 x)-A_l \chi_l'(m_2 x)\right]}{\xi_l'(x)\left[\psi_l(m_2 x)-A_l \chi_l(m_2 x)\right]- m_2\xi_l(x)\left[\psi_l'(m_2 x)-A_l \chi_l'(m_2 x)\right]},
\end{align}
\end{subequations}
\end{widetext}
where, $x = k_o a_2$, $\quad x_1 = k_o a_1$, $\quad m_1 = \frac{n_1}{n_b}$, $\quad m_2 = \frac{n_2}{n_b}$, the Riccati-Bessel and Neumann functions are $\psi_l(z)=z j_l(z)$, $\chi_l (z) = -zy_l(z)$, respectively and the Riccatti-Hankel function is $\quad \xi_l(z)=z h_l^{(1)}(z)$ = $\psi_l(z)+i\chi_l(z)$, and $A_l$ is the auxiliary coefficient, defined as:
\begin{align}
A_l =
\frac{m_2 \psi_l(m_2 x_1)\psi_l'(m_1 x_1)-
m_1 \psi_l'(m_2 x_1)\psi_l(m_1 x_1)}{m_2 \psi_l(m_1 x_1)\chi_l'(m_2 x_1) -m_1 \psi_l'(m_1 x_1)\chi_l(m_2 x_1)}
\end{align}
The coefficients $a_{\nu}$, $\alpha_{\nu}$, $\beta_{\nu}$, and $b_{\nu}$ represent the amplitudes of the allowed radial solutions in the core, shell, and surrounding medium. The presence of the shell introduces an additional scattering channel, modifying the resonance condition and redistributing the modal fields.

The operators($R^{(i)}$ and $S^{(i)}$) remain diagonal in the VSH basis, but their diagonal elements now incorporate the layered structure through the core-shell Mie coefficients. These modified operators form the basis for describing coupling and hybridization in photonic molecule systems.

\section{VSH Translation Coefficients}
When expressing the scattered field of sphere $j$ in the coordinate system of sphere $i$, VSH translation theorem has been applied. For a separation distance ($d = |R_i - R_j|$), the outgoing VSH around sphere $j$ satisfies:

\begin{align} \label{Eq:translation-appendix}
  E^{(j)}_{\nu'}(k r_j)=\sum_{\nu}
\mathcal{T}^{(j\rightarrow i)}_{\nu\nu'}(\omega,d)E^{(i)}_{\nu}(k r_i), 
\end{align}
where, $\mathcal{T}^{(j\rightarrow i)}_{\nu\nu'}(\omega,d)$ constitute the translation operator using bisphere Hamiltonian. The translation operator can be written compactly as:

\begin{widetext}
    \begin{align}  \label{Eq:Wigner_3j}
\mathcal{T}^{(j\rightarrow i)}_{lm,l'm'}(k,d)=\sum_{L}(-1)^{m'}i^{l'-l}(2L+1)
\begin{pmatrix}
l & l' & L \\
m & -m' & m'-m
\end{pmatrix}
\begin{pmatrix}
l & l' & L \\
\beta & \beta' & -\beta
\end{pmatrix}
h^{(1)}_{L}(k d),
\end{align}
\end{widetext}

where the Wigner 3-j symbols determine the angular coupling, the spherical Hankel function $h^{(1)}_{L}(k d)$ provides the radial dependence, and the indices ($\beta,\beta'\in{M,N}$) distinguish TE/TM VSHs.

The equation can then be written, for simplicity, as:
\begin{align}
\mathcal{T}^{(j\rightarrow i)}(\omega,d)\equiv\left[\mathcal{T}^{(j\rightarrow i)}_{\nu\nu'}(\omega,d)\right]_{\nu,\nu'}
 \end{align}

This equation expresses the VSH associated with sphere $j$ in the coordinate system of sphere $i$. The translation operator $\mathcal{T}^{(j\rightarrow i)}_{\nu\nu'}$ contains the full angular and radial structure of this re-expansion. The two Wigner 3-j symbols encode the rules of angular momentum coupling that are required to rotate a multipole ($l',m',\beta'$) at $R_j$ into a superposition of multipoles ($l,m,\beta$) about $R_i$. These terms arise directly from the Clebsch–Gordan algebra of spherical harmonics and determine the mixing of different azimuthal and polar components \cite{CRUZAN1962, Waterman1971, Miyazaki2000}. The spherical Hankel function $h_L^{(1)}(kd)$ provides the radial dependence of the translation and accounts for the propagation of the scattered field over the center-to-center separation '$d$'. 

\section{Bisphere single mode tight-binding Hamiltonian}
As ($\mathbf{f}_{\nu}$) contains the projections of the external field onto the VSH basis, it vanishes for naturally occurring resonances. Hence, Eq.~\ref{Eq:Matrix form of chracteristic eq} reduces to the homogeneous system
\begin{align}
\big[\mathbf{I}_\nu - \mathbf{V}_{\nu}(\omega,d)\big] \mathbf{b}_{\nu} = 0.
\end{align}

A non-trivial solution for the scattering coefficients ($\mathbf{b}_{\nu}\neq 0$) exists only when the corresponding operator is singular, which yields the characteristic equation of the coupled system:
\begin{align}
\det\big[\mathbf{I}_\nu - \mathbf{V}_{\nu}(\omega,d)\big] = 0.
\label{Eq:secular_full}
\end{align}

Using the block structure of the interaction operator ($\mathbf{V}_\nu$), the determinant can be reduced to
\begin{align}
\det\big[\mathbf{I}_\nu - \mathcal{V}^{21}(\omega,d)\mathcal{V}^{12}(\omega,d)\big] = 0.
\label{Eq:secular_reduced}
\end{align}

To obtain an analytically tractable form, we project Eq.~\ref{Eq:secular_reduced} onto a dominant resonant multipole ($\nu_0$), which is justified for spectrally isolated WGM resonances. Under this single-mode approximation, the matrix equation reduces to a scalar condition:
\begin{align}
1 - \mathcal{V}^{21}_{\nu_0\nu_0}(\omega,d)\mathcal{V}^{12}_{\nu_0\nu_0}(\omega,d) = 0.
\end{align}

\begin{figure*} [htbp]
    \centering
    \includegraphics[scale=0.25]{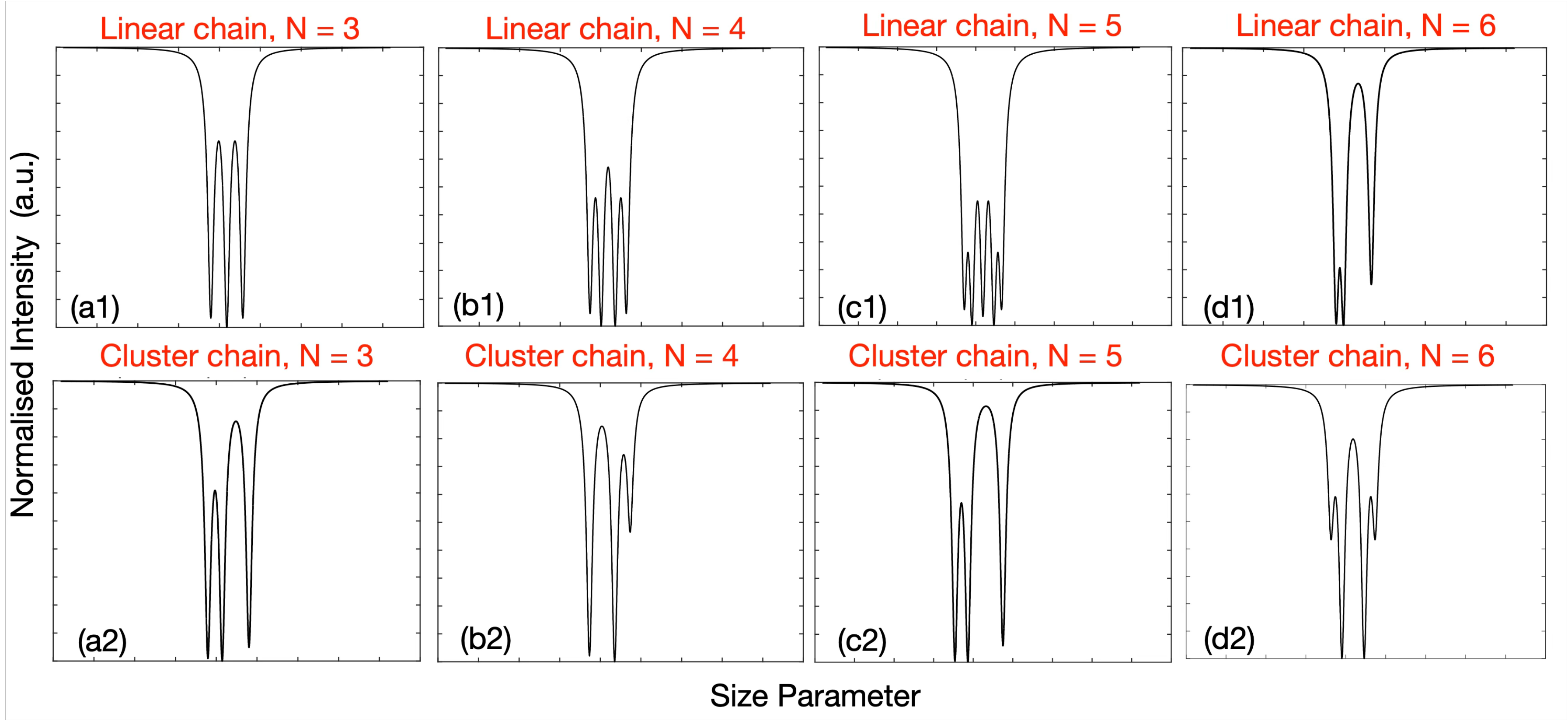}
    \caption{SMTB description of collective supermode formation in multi-resonator photonic molecules. The corresponding nearest-neighbour Hamiltonians for the linear-chain (a1-d1) and compact topologies (a2-d2) are constructed from the isolated parent WGM resonance and effective inter-resonator coupling, and their eigenvalues show the topology-dependent splitting and degeneracy of the resulting collective modes.}
    \label{fig:A9}
\end{figure*}

Substituting the definitions of the coupling operators (Eq.~\ref{Eq:Definition mathcal V}), and using the diagonal nature of the Mie scattering operator for spherical resonators, we obtain
\begin{align}
1 - S^{(1)}_{\nu_0}(\omega)S^{(2)}_{\nu_0}(\omega)\mathcal{T}_{\nu_0}^2(\omega,d) = 0,
\label{Eq:final_secular_scalar}
\end{align}
For two identical spheres ($S^{(1)}_{\nu_0}(\omega) = S^{(2)}_{\nu_0}(\omega) \equiv S_{\nu_0}(\omega)$), the Eq.~\ref{Eq:final_secular_scalar} simplifies to:
\begin{align}
1 - S_{\nu_0}^2(\omega)\mathcal{T}_{\nu_0}^2(\omega,d) = 0,
\end{align}

Expressing the Mie coefficients in its exact form,
\begin{align} \label{Eq:numerator/denominator}
S_{\nu_0}(\omega) = \frac{N_{\nu_0}(\omega)}{D_{\nu_0}(\omega)},
\end{align}
where, $N_{\nu_0}(\omega)$ in the numerator of the scattering coefficients and $D_{\nu_0}(\omega)$ is the denominator of scattering coefficients, as obtained in Eq.\ref{eq:Mie scattering in prsitine spheres} and Eq.\ref{eq:Mie scattering in core-shell spheres}. Further substituting these into Eq.~\ref{Eq:final_secular_scalar}, we obtain
\begin{align}
D_{\nu_0}(\omega) = \pm N_{\nu_0}(\omega)\mathcal{T}_{\nu_0}(\omega,d).
\label{Eq:eigen_eq_exact}
\end{align}

The natural resonance of an isolated single sphere at $\omega_0$ is defined by $D_{\nu_0}(\omega_0) = 0$, and expanding the above equation Eq.~\ref{Eq:eigen_eq_exact} to first order around ($\omega_0$) :
\begin{align}
D_{\nu_0}(\omega) \approx D'_{\nu_0}(\omega_0)(\omega - \omega_0),
\end{align}
and evaluating the slowly varying terms at ($\omega_0$), we obtain
\begin{align}
D'_{\nu_0}(\omega_0)(\omega - \omega_0) = \pm N_{\nu_0}(\omega_0)\mathcal{T}_{\nu_0}(\omega_0,d).
\end{align}

Solving for the eigenfrequencies yields the split modes:
\begin{align}
\omega_{\pm} = \omega_0 \pm g(d)
\end{align}
where the coupling strength is given by
\begin{align}
g(d) = \frac{N_{\nu_0}(\omega_0)}{D'_{\nu_0}(\omega_0)}\mathcal{T}_{\nu_0}(\omega_0,d).
\end{align}

Accordingly, the frequency splitting between the bonding and antibonding modes is given by:
\begin{align}
\Delta \omega = \omega_{+} - \omega_{-} = 2g(d).
\end{align}

\section{Multisphere single mode tight-binding Hamiltonian}
For larger photonic molecules, the same pairwise coupling framework is generalized to an \(N\times N\) Hamiltonian. The Hamiltonian is defined as:
$$ H_{ij}= \begin{cases} \omega_0, & i=j,\\ g, & \text{if resonators }i\text{ and }j\text{ are coupled},\\ 0, & \text{otherwise}, \end{cases} $$

where \(\omega_0\) is the resonance frequency of an isolated resonator and \(g\) is the coupling parameter between coupled resonators. The molecular topology is incorporated through the positions of the nonzero off-diagonal coupling terms in the Hamiltonian. The eigenvalues of \(H\) represent the collective supermode frequencies of the photonic molecule. Thus, different molecular topologies result in distinct supermode spectra through their respective coupling connectivity. Figure \ref{fig:A9} shows the corresponding Hamiltonian matrices and calculated eigenvalue spectra for the linear and clustered multi-resonator photonic molecule configurations.

\bibliography{main}

\end{document}